\documentclass[11pt,a4paper]{article}
\usepackage{iftex}
\ifPDFTeX\pdfoutput=1\fi
\usepackage{jheppub}
\usepackage[T1]{fontenc}
\ifPDFTeX\usepackage[utf8]{inputenc}\fi
\usepackage{amsmath,amssymb,mathtools,bm}
\usepackage{braket,booktabs,graphicx,array,comment,hyperref}
\usepackage{placeins}
\usepackage{microtype}
\usepackage{tikz}
\usetikzlibrary{decorations.markings}
\graphicspath{{figures/}{./}}
\allowdisplaybreaks
\newcommand{\Tr}{\operatorname{Tr}}
\newcommand{\diag}{\operatorname{diag}}

\newcommand{\ACP}{A^{\mathrm{CP}}}
\newcommand{\DACP}{\Delta A^{\mathrm{CP}}}
\newcommand{\dd}{\mathrm{d}}
\newcommand{\ii}{\mathrm{i}}
\newcommand{\Liouv}{\mathcal{L}}
\newcommand{\CP}{\mathrm{CP}}
\newcommand{\daughter}{\mathrm{daughter}}
\newcommand{\parent}{\mathrm{parent}}
\newcommand{\st}{\mathrm{st}}
\newcommand{\vis}{\mathrm{vis}}
\newcommand{\CX}{\mathrm{CX}}

\title{CP asymmetry and visible decay in \(3+1\) neutrino oscillations on a quantum computer}
\author[a,b]{Amartya Sengupta}
\author[c]{Sidhartha Samtani}
\author[a]{Ani Girgvliani}
\author[a]{Dejan Stojkovic}
\affiliation[a]{Department of Physics, The State University of New York, Buffalo, New York 14260, USA}
\affiliation[b]{Fermi National Accelerator Laboratory,
Batavia, Illinois 60510, USA}
\affiliation[c]{Homer L. Dodge Department of Physics and Astronomy,
University of Oklahoma, Norman, Oklahoma 73019, USA}
\emailAdd{amartyas@buffalo.edu}
\emailAdd{sidhartha.samtani@ou.edu}
\emailAdd{anigirgv@buffalo.edu}
\emailAdd{ds77@buffalo.edu}

\abstract{
We use quantum simulation to study how visible neutrino decay modifies the vacuum CP asymmetry in muon-to-electron oscillations within a $3+1$ neutrino scenario. The decay channel preserves CP symmetry, with CP violation arising from phases in the mixing matrix. We track the energy redistribution from visible decay and distinguish how parent attenuation and daughter regeneration modify the CP asymmetry. For the reference benchmark, the regenerated contribution is $4.85\times10^{-4}$, approximately $46\%$ of the total decay-induced shift. The full circuit for an effective two-energy model agrees with independent classical evolution. On an IBM quantum processor, we prepare the daughter state conditioned on decay and restore its physical normalization using fixed source and decay probabilities. After correction for measurement errors, the flavor-probability and coherence measurements both agree with the prediction within one standard error. Null controls test the suppression of the regenerated signal when daughter coherence or the relevant CP-sensitive mixing factor vanishes.
}

\begin{document}
\maketitle

\section{Introduction}
\label{sec:introduction}

Visible neutrino decay modifies oscillation probabilities through parent attenuation and the regeneration of detectable daughter neutrinos \cite{Kopp:2026tnx,Lindner:2001fx,Gago:2017zzy,Coloma:2017zpg}.
In vacuum, the CP asymmetry depends on complex mixing invariants and relative propagation phases \cite{Jarlskog:2004be,Cabibbo:1977nk,Barger:1980jm,PhysRevD.18.958}. Neutrino decay can therefore modify an existing asymmetry even when the decay interaction itself respects CP. The decay of a mass eigenstate reduces its interference with the surviving states, while its
daughter neutrinos contribute to the appearance signal at lower energies. If the unobserved decay products do not distinguish the daughter masses, the daughter amplitudes can also interfere at detection \cite{Lindner:2001fx,Kopp:2026tnx}.
In this paper we examine how parent attenuation and daughter interference
contribute separately to the decay-induced change in the CP asymmetry.

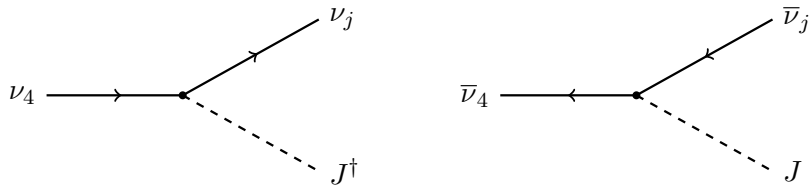
\begin{figure}[tbp]
\centering
\begin{tikzpicture}[
    line width=0.85pt,
    fermion/.style={
        postaction={decorate},
        decoration={
            markings,
            mark=at position 0.55 with {\arrow{>}}
        }
    },
    antifermion/.style={
        postaction={decorate},
        decoration={
            markings,
            mark=at position 0.55 with {\arrow{<}}
        }
    },
    scalar/.style={dashed}
]
    \coordinate (v) at (1.8,0);
    \draw[fermion] (0,0)
        node[left] {$\nu_4$} -- (v);
    \draw[fermion] (v) -- (3.6,1.0)
        node[right] {$\nu_j$};
    \draw[scalar] (v) -- (3.6,-1.0)
        node[right] {$J^\dagger$};
    \fill (v) circle (1.5pt);

    \begin{scope}[xshift=6cm]
        \coordinate (vb) at (1.8,0);
        \draw[antifermion] (0,0)
            node[left] {$\overline{\nu}_4$} -- (vb);
        \draw[antifermion] (vb) -- (3.6,1.0)
            node[right] {$\overline{\nu}_j$};
        \draw[scalar] (vb) -- (3.6,-1.0)
            node[right] {$J$};
        \fill (vb) circle (1.5pt);
    \end{scope}
\end{tikzpicture}
\caption{
Lepton-number-conserving decay of the unstable Dirac state
$\nu_4$ and its CP-conjugate process. Solid lines denote neutrinos, with arrows indicating fermion-number
flow; dashed lines denote the unobserved scalar.
The daughter index is $j=1,2,3$.
The vertices represent the effective decay interaction.
}
\label{fig:decay-channel}
\end{figure}

We address this question in a $3+1$ scenario containing three
active flavors and one sterile flavor. The additional mixing phases enrich the CP dependence of appearance probabilities \cite{Fiza:2021gvq}, while an unstable fourth mass eigenstate connects this dependence to decay. Sterile-neutrino decay has been investigated in atmospheric and short-baseline neutrino experiments
\cite{Moss:2017pur,Dentler:2019dhz,Dasgupta:2021ies}, including its interplay with CP phases \cite{Becerra-Garcia:2021gcd}. We consider the lepton-number-conserving Dirac channels
\begin{equation}
 \nu_4 \longrightarrow \nu_j+J^\dagger,
 \qquad
 \overline{\nu}_4 \longrightarrow \overline{\nu}_j+J,
 \qquad j=1,2,3,
 \label{eq:intro-decay}
\end{equation}
where $\nu_4$ is the unstable mass eigenstate, $\nu_j$ denotes a lighter daughter, and $J$ is a very light scalar.
Interactions between neutrinos and a light scalar can induce these decays
\cite{Chikashige:1980ui,GELMINI1981411,Kim:1990km,Abdullahi:2020rge}. Figure~\ref{fig:decay-channel} illustrates the two processes.
We impose CP symmetry on the effective decay channel and retain the CP-violating phases in the neutrino mixing matrix.

The regenerated CP contribution depends on whether the unobserved decay products distinguish the daughter mass states. When these alternatives remain indistinguishable, the daughter amplitudes can interfere at detection
\cite{Lindner:2001fx,Kopp:2026tnx}.
This interference is suppressed when the unobserved decay products
carry information that distinguishes the daughter mass eigenstates
\cite{Englert:1996zz,Akhmedov:2009rb,Akhmedov:2010ua,Kayser:1981ye,Giunti:1991ca,Kiers:1995zj}.
Coherence alone does not establish a regenerated CP asymmetry. The interference must also contain a CP-odd mixing factor and a nonvanishing propagation-phase contribution. We vary the daughter coherence at fixed regenerated populations to isolate its contribution to the appearance asymmetry.

Open-system methods provide a natural description of this
evolution~\cite{Breuer:2007juk,Caban:2005ue}. Earlier studies treated dissipative neutrino oscillations \cite{Benatti:2000ph,Loreti:1994ry,Coelho:2017byq,DeRomeri:2023dht,Stankevich:2020sja,Purtova:2023hcy,Stankevich:2023qdi,dosSantos:2023skk,Gago:2000qc,Lisi:2000zt,Stankevich:2020icp} and formulated particle decay on a state space that includes the decay products \cite{Bertlmann:2006fn,Stankevich:2024xyc}.
Kopp and Parker constructed an energy-resolved description of
visible decay that retains unresolved daughter interference and decay cascades \cite{Kopp:2026tnx}.
Building on this framework, we derive the decay-induced change in the vacuum $\nu_\mu\to\nu_e$ CP asymmetry relative to stable $3+1$ propagation. We separate this change into parent attenuation and daughter regeneration, and identify the conditions under which the regenerated term vanishes.
The resulting controls distinguish a coherent daughter state
from a state whose interference contributes to the CP observable.

This relation between coherence and appearance motivates a
quantum-computer experiment. Previous implementations have studied neutrino mixing and oscillations
\cite{Arguelles:2019phs,Molewski:2021ogs}, including
CP-dependent three-flavor evolution
\cite{Nguyen:2022snr,Singh:2024vpu}.
Related studies have explored open-system quantum walks
\cite{Sahu:2023csa}, collective oscillations on
superconducting and trapped-ion devices
\cite{Hall:2021rbv,Amitrano:2022yyn},
and three-flavor qubit and qutrit encodings
\cite{Turro:2024shh,Spagnoli:2025etu}.
Symmetry-based constructions have further reduced the quantum
registers required for collective models \cite{Bleau:2026iiq}.
Our experiment focuses on the regenerated CP contribution:
we control the daughter coherence and test whether flavor
probabilities and a separate coherence measurement reconstruct the same physical signal.

We construct a five-qubit circuit for the complete decay channel of an effective two-energy model, including parent survival, energy transfer, and daughter interference. Its noiseless simulation agrees with independent classical evolution. On the IBM processor, we prepare the
normalized daughter state with a two-qubit circuit and reconstruct the regenerated contribution using source and decay weights fixed by the reference benchmark.

We choose the two-energy benchmark so that exact classical evolution provides a reference for the processor measurements. The experiment tests the controlled preparation of daughter coherence and the measurement of its contribution to the CP asymmetry. Its purpose is to assess the accuracy of these operations on a quantum processor; the benchmark itself is readily tractable classically.
Appendix~\ref{app:circuit} develops the full channel construction, including parent survival and daughter regeneration. This construction provides a starting point for incorporating visible decay into quantum simulations of interacting neutrino systems, where the computational cost of classical evolution can become substantial.

The predicted regenerated asymmetry is $4.85\times10^{-4}$,
approximately $46\%$ of the total decay-induced shift. After correction for measurement assignment errors\footnote{An assignment error occurs when one computational-basis outcome is recorded as another.}, the flavor-probability and coherence measurements both agree with the
prediction within one standard error. Six predefined null controls, in which coherence or the relevant CP-sensitive mixing factor vanishes, are consistent with zero within two standard errors.

Section~\ref{sec:physics} develops the decay channel and CP
decomposition, section~\ref{sec:quantum} presents the benchmark and circuits, and section~\ref{sec:results} reports the numerical and processor results. Section~\ref{sec:discussion} discusses their implications and future extensions. The appendices provide the spectral input, circuit constructions, validation, calibration, and statistical analysis.

\section{Visible decay and the CP asymmetry}
\label{sec:physics}

We study vacuum propagation of four Dirac neutrinos with one unstable
mass eigenstate, $\nu_4$. The flavor indices
$\alpha,\beta\in\{e,\mu,\tau,s\}$ include the sterile flavor $s$, and
$i\in\{1,2,3,4\}$ labels the mass eigenstates. The two bases are related by a unitary matrix $U$
\cite{1962PThPh..28..870M,Fiza:2021gvq,Pontecorvo:1957qd,Pontecorvo:1967fh},
\begin{equation}
 \ket{\nu_\alpha}=\sum_iU_{\alpha i}^{*}\ket{\nu_i},
 \qquad
 \ket{\bar\nu_\alpha}=\sum_iU_{\alpha i}\ket{\bar\nu_i}.
 \label{eq:flavor-states}
\end{equation}
We fix the rotation order as
\begin{equation}
 U=R_{34}\widetilde R_{24}\widetilde R_{14}
 R_{23}\widetilde R_{13}R_{12}.
 \label{eq:mixing}
\end{equation}
The matrix $R_{ij}$ contains the angle $\theta_{ij}$, while
$\widetilde R_{ij}$ also contains the phase $\delta_{ij}$.
Their entries are specified in appendix~\ref{app:conventions}. The phase vector is
$\bm\delta=(\delta_{13},\delta_{14},\delta_{24})$.
These phases enter the appearance amplitudes through interference
between mass eigenstates~\cite{Fiza:2021gvq}.

For the decay channels in eq.~\eqref{eq:intro-decay}, we impose CP symmetry on the effective decay dynamics and take the transition coefficients to be real in the mixing convention of eq.~\eqref{eq:mixing}. All CP-odd phases then enter through $U$. More general treatments of decay-induced nonunitary neutrino propagation allow additional CP-violating parameters in the effective Hamiltonian
\cite{Berryman:2014yoa}.

All probabilities refer to a single neutrino produced at the source. We use natural units, $\hbar=c=1$, and identify propagation time with the baseline $L$ in the ultrarelativistic limit.

\subsection{Energy transfer and the reduced neutrino state}
\label{subsec:channel}

We use the basis $\ket{i,n}=\ket{\nu_i,E_n}$, where $E_n$ is the
representative energy of bin $n$. The density-matrix block
$\rho^{(n)}$ retains coherence between masses at that energy.
Averaging over macroscopically separated energies removes the
off-diagonal energy blocks, giving
$\rho=\bigoplus_n\rho^{(n)}$~\cite{Kopp:2026tnx}.
The vacuum Hamiltonian~\cite{Akhmedov:2009rb} is
\begin{equation}
 H=\sum_{i,n}\omega_{in}\ket{i,n}\!\bra{i,n},
 \qquad
 \omega_{in}=\frac{\Delta m_{i1}^{2}}{2E_n},
 \qquad
 \Delta m_{i1}^{2}=m_i^2-m_1^2,
 \label{eq:hamiltonian}
\end{equation}
with $m_i$ the mass of $\nu_i$.
The decay width per unit propagation length is
$\Gamma_{4n}=\Gamma_4(E_n)=\alpha_4/E_n$, where
$\alpha_4=m_4/\tau_4$ and $\tau_4$ is the rest-frame lifetime.

In the decay expressions, $j,k\in\{1,2,3\}$ label daughter
mass eigenstates, and $E'$ denotes the daughter energy.
The partial width $\Gamma_{4j}(E_n)$ describes decay into
$\nu_j$, integrated over all allowed daughter energies.
The differential partial width
$\eta_{4j}(E_n,E')=\dd\Gamma_{4j}(E_n,E')/\dd E'$
determines the decay rate into each daughter-energy bin.
For the transition $\ket{4,n}\to\ket{j,m}$, the corresponding
jump amplitude $\gamma_{4j}^{(nm)}$ satisfies \cite{Kopp:2026tnx}
\begin{equation}
 \left|\gamma_{4j}^{(nm)}\right|^2
 =\int_{E_m^-}^{E_m^+}\eta_{4j}(E_n,E')\,\dd E',
 \qquad
 \Gamma_{4n}=\sum_{m,j}\left|\gamma_{4j}^{(nm)}\right|^2.
 \label{eq:bin-rates}
\end{equation}
Here $E_m^-$ and $E_m^+$ denote the edges of daughter-energy
bin $m$, and $\Gamma_{4n}$ is the total decay rate of
$\nu_4$ at energy $E_n$.
The full width is exhausted by the represented decay modes in the two-energy benchmark. The sum includes all decay channels and daughter energies retained
in the model, so that regeneration preserves the total probability. \cite{Gago:2017zzy,Porto-Silva:2020gma}.

For the pair $4\to1,3$, we introduce a real overlap
$\zeta\in[0,1]$ between the environmental states associated with the
two daughter alternatives. The Lindblad jump operators are
\begin{align}
 L_a^{(nm)}&=\gamma_{41}^{(nm)}\ket{1,m}\!\bra{4,n}
       +\zeta\gamma_{43}^{(nm)}\ket{3,m}\!\bra{4,n},\nonumber\\
 L_b^{(nm)}&=\sqrt{1-\zeta^2}\,
       \gamma_{43}^{(nm)}\ket{3,m}\!\bra{4,n}.
 \label{eq:overlap-jumps}
\end{align}
The normalized environmental state $\ket{e_j}$ is associated
with daughter $\nu_j$. For the two alternatives considered here,
$\langle e_3|e_1\rangle=\zeta$.
At $\zeta=1$ they are identical and do not distinguish the daughter
masses; at $\zeta=0$ they are orthogonal and distinguish them
completely~\cite{Englert:1996zz}. We treat $\zeta$ as an independent
parameter of the effective channel and assign the same value to the
CP-conjugate processes.

The open-system generator has the Gorini--Kossakowski--Sudarshan--Lindblad
form~\cite{Gorini:1975nb,Lindblad:1975ef,Brasil:2012trs},
with energy-dependent jumps as in refs.~\cite{Stankevich:2024xyc,Kopp:2026tnx}:
\begin{equation}
 \frac{\dd\rho}{\dd L}
 =-\ii[H,\rho]
 +\sum_{n,m}\sum_{r=a,b}
 \left(L_r^{(nm)}\rho L_r^{(nm)\dagger}
 -\frac12\{L_r^{(nm)\dagger}L_r^{(nm)},\rho\}\right)
 \equiv\Liouv_\zeta(\rho).
 \label{eq:master}
\end{equation}
The anticommutator is $\{A,B\}=AB+BA$.
The generated evolution is completely positive and trace preserving.
The gain term populates the lower-energy daughter states, while the
anticommutator term accounts for the loss of parent population.
At fixed partial widths,
\begin{equation}
 \sum_{r=a,b}L_r^{(nm)\dagger}L_r^{(nm)}
 =\left(\left|\gamma_{41}^{(nm)}\right|^2
       +\left|\gamma_{43}^{(nm)}\right|^2\right)
       \ket{4,n}\!\bra{4,n}.
 \label{eq:fixed-loss}
\end{equation}
The loss term therefore remains fixed throughout the overlap scan.
Only the off-diagonal daughter source changes, in proportion to
$\zeta$. This construction separates coherence from the amount of
probability transferred by decay.

\subsection{CP asymmetry and daughter coherence}
\label{subsec:cp}

For an initial flavor $\alpha$ in bin $n$, propagation without a decay
has amplitude
\begin{equation}
 \mathcal A_{\alpha\beta}^{\nu,0}
 =\sum_i U_{\beta i}U_{\alpha i}^{*}
 e^{-\ii\Phi_{in}}e^{-\Gamma_{in}L/2},
 \qquad
 \Phi_{in}=\omega_{in}L,\qquad
 \Gamma_{in}=\Gamma_{4n}\delta_{i4}.
 \label{eq:parent-amplitude}
\end{equation}
The symbol $\delta_{i4}$ denotes the Kronecker delta.
The stable amplitude is recovered at $\Gamma_{4n}=0$.
Complex conjugation of the mixing matrix $U$ gives the antineutrino
amplitude, while the propagation phases remain unchanged
\cite{Bilenky:1976yj,Bilenky:1978nj,Strumia:2006db}.

The source population of the unstable mass state is
$p_{\alpha4}^{\mathrm{in}}=|U_{\alpha4}|^2$.
Integrating the gain term over the decay position, as in the regeneration
treatment of refs.~\cite{Lindner:2001fx,Abdullahi:2020rge},
we obtain the daughter density matrix
\begin{align}
 \rho_{\alpha,jk}^{(m),\daughter}(L)
 &=p_{\alpha4}^{\mathrm{in}}\,
   \gamma_{4j}^{(nm)}\gamma_{4k}^{(nm)*}
   S_{jk}^{(nm)} I_{jk}^{(nm)}(L),
 \label{eq:daughter-density}\\
 I_{jk}^{(nm)}(L)
 &=\int_0^L\dd\ell\,e^{-\Gamma_{4n}\ell}
    e^{-\ii\Delta\omega_{jk}^{(m)}(L-\ell)}
 =\frac{e^{-\ii\Delta\omega_{jk}^{(m)}L}-e^{-\Gamma_{4n}L}}
        {\Gamma_{4n}-\ii\Delta\omega_{jk}^{(m)}},
 \label{eq:integral}
\end{align}
where $\ell$ is the decay position,
$\Delta\omega_{jk}^{(m)}=\omega_{jm}-\omega_{km}$,
and $S_{jk}^{(nm)}=\langle e_k^{(nm)}|e_j^{(nm)}\rangle$
is the environmental overlap matrix.
For eq.~\eqref{eq:overlap-jumps}, $S_{11}=S_{33}=1$ and
$S_{13}=S_{31}=\zeta$; the energy indices are implicit in these entries.
The gain term sums contributions from different decay positions
incoherently, while retaining interference between daughter masses
produced at each position~\cite{Lindner:2001fx,Kopp:2026tnx}.
The diagonal kernel is
$I_{jj}=(1-e^{-\Gamma_{4n}L})/\Gamma_{4n}$.

Projecting the parent and daughter contributions onto flavor
$\beta$, we obtain the appearance probability
\begin{equation}
 P^\nu_{\alpha\beta}(m\leftarrow n;L)
 =\delta_{mn}|\mathcal A_{\alpha\beta}^{\nu,0}|^2
 +\sum_{j,k}U_{\beta j}U_{\beta k}^{*}
       \rho_{\alpha,jk}^{(m),\daughter}(L).
 \label{eq:probability}
\end{equation}
The undecayed and regenerated sectors correspond to orthogonal
environmental states, with no emitted scalar and an emitted scalar,
respectively. Their probabilities therefore add without a parent--daughter
interference term~\cite{Lindner:2001fx,Abdullahi:2020rge}.
We define the CP asymmetry as the difference between the
appearance probabilities for $\nu_\alpha\to\nu_\beta$ and
$\bar\nu_\alpha\to\bar\nu_\beta$ at the same energies and baseline,
\begin{equation}
 \ACP_{\alpha\beta}(m\leftarrow n;L)
 =P^\nu_{\alpha\beta}(m\leftarrow n;L)
 -P^{\bar\nu}_{\alpha\beta}(m\leftarrow n;L).
 \label{eq:cp-asymmetry}
\end{equation}
Summing over the final-energy bins, we define the decay-induced change relative to stable propagation as\footnote{The labels $\vis$ and $\st$ denote visible decay and stable
propagation, respectively.}
\begin{equation}
 \DACP_{\alpha\beta}(n;L)
 =\sum_m\ACP_{\alpha\beta,\vis}(m\leftarrow n;L)
 -\ACP_{\alpha\beta,\st}(E_n,L).
 \label{eq:cp-definitions}
\end{equation}

The rephasing-invariant mixing quartets
$J_{\alpha\beta}^{ij}=U_{\beta i}U_{\alpha i}^{*}
U_{\beta j}^{*}U_{\alpha j}$ enter the oscillation probabilities.
Their imaginary parts are Jarlskog-type invariants that determine
the CP-odd contribution to stable oscillations
\cite{Jarlskog:2004be,Jarlskog:1985ht}:
\begin{equation}
 \ACP_{\alpha\beta,\st}
 =4\sum_{i>j}\operatorname{Im}J_{\alpha\beta}^{ij}
                 \sin(\Phi_{in}-\Phi_{jn}).
 \label{eq:stable-cp}
\end{equation}
Attenuation modifies the parent CP asymmetry through interference
terms involving the unstable mass eigenstate $\nu_4$. The decay-induced
change separates into parent and daughter contributions,
\begin{equation}
 \DACP_{\alpha\beta}
 =\DACP_{\alpha\beta,\parent}
 +\sum_m\ACP_{\alpha\beta,\daughter}(m\leftarrow n;L).
 \label{eq:cp-decomposition}
\end{equation}
Writing $\kappa_n=\Gamma_{4n}L$, we obtain
\begin{align}
 \DACP_{\alpha\beta,\parent}
 &=4\bigl(e^{-\kappa_n/2}-1\bigr)
 \sum_{j<4}\operatorname{Im}J_{\alpha\beta}^{4j}
 \sin(\Phi_{4n}-\Phi_{jn}),
 \label{eq:parent-shift}\\
 \ACP_{\alpha\beta,\daughter}(m\leftarrow n;L)
 &=-4p_{\alpha4}^{\mathrm{in}}
 \sum_{j<k}\operatorname{Im}Z_{\beta;jk}^{(nm)}
 \operatorname{Im}I_{jk}^{(nm)}.
 \label{eq:daughter-cp}
\end{align}
The factor
\begin{equation}
 Z_{\beta;jk}^{(nm)}
 =U_{\beta j}U_{\beta k}^{*}
 \gamma_{4j}^{(nm)}\gamma_{4k}^{(nm)*}S_{jk}^{(nm)}
 \label{eq:Z-invariant}
\end{equation}
combines the flavor projection, decay amplitudes, and environmental
overlap for each daughter pair.\footnote{Equation~\eqref{eq:daughter-cp}
uses real overlaps and the CP-conjugation rule
$\bar\gamma_{4j}^{(nm)}=\gamma_{4j}^{(nm)*}$.}

Only the mixing quartets $J_{\alpha\beta}^{4j}$ with $j<4$ enter
the parent contribution. The regenerated contribution instead carries the initial
population $p_{\alpha4}^{\mathrm{in}}$ and depends on interference
between the lighter daughters. Its CP-odd mixing and decay factor
is multiplied by the imaginary propagation kernel, which accounts
for the relative phase accumulated after decay and its averaging
over the decay position.

For the $4\to1,2$ control, the overlap matrix has
$S_{11}=S_{22}=1$ and $S_{12}=S_{21}=\zeta$.
For $\alpha=\mu$ and $\beta=e$, the entries $U_{e1}$ and $U_{e2}$ are
real in our convention. Real $4\to1,2$ amplitudes therefore produce
$\operatorname{Im}Z_{e;12}^{(nm)}=0$ even at $\zeta=1$.
A pair containing $\nu_3$ instead probes
$U_{e3}\propto e^{-\ii\delta_{13}}$.
Thus $4\to1,2$ supplies a coherent null channel for the regenerated
CP observable, while $4\to1,3$ carries its signal.
The phases $\delta_{14}$ and $\delta_{24}$ remain in the parent
quartets. For either daughter pair the complete asymmetry satisfies
\begin{equation}
 \ACP_{\alpha\beta}(\bm0)=0,
 \qquad
 \ACP_{\alpha\beta}(-\bm\delta)
 =-\ACP_{\alpha\beta}(\bm\delta).
 \label{eq:cp-controls}
\end{equation}

For the circuit construction, we specialize to an effective two-energy
model with source energy $E_H$ and daughter energy $E_L$. We define
\begin{equation}
 \kappa=\Gamma_4(E_H)L,\qquad D=1-e^{-\kappa},\qquad
 \phi_{31}=\frac{\Delta m_{31}^2L}{2E_L},\qquad
 F_{13}=\frac{\kappa(e^{\ii\phi_{31}}-e^{-\kappa})}
                   {\kappa+\ii\phi_{31}}.
 \label{eq:dimensionless-kernel}
\end{equation}
Here $\kappa$ is the dimensionless integrated decay rate, and
$\phi_{31}$ is the relative propagation phase between $\nu_3$
and $\nu_1$ over the baseline $L$ at energy $E_L$.
$D$ is the decay probability of an initial $\nu_4$ and
$F_{13}=\Gamma_4(E_H)I_{13}$ is the dimensionless interference kernel.
The branching fraction for decay into $\nu_j$ is
$B_j=\Gamma_{4j}(E_H)/\Gamma_4(E_H)$.
For the two channels $4\to1$ and $4\to3$, these fractions satisfy
$B_1+B_3=1$. For a muon-flavor source, the surviving parent population $p_4$, the
daughter populations $p_j$, and the daughter coherence $\rho_{13}^L$ are
\begin{equation}
 p_4=p_{\mu4}^{\mathrm{in}}e^{-\kappa},\qquad
 p_j=p_{\mu4}^{\mathrm{in}}B_jD,\qquad
 \rho_{13}^L=\zeta p_{\mu4}^{\mathrm{in}}\sqrt{B_1B_3}\,F_{13}.
 \label{eq:two-level-state}
\end{equation}
The superscript $L$ and the states $\ket{i,L}$ refer to the daughter energy $E_L$.
Here and below $p_j$ refers only to regenerated population at $E_L$;
the original light components remain in the parent-energy sector.

For an operator $O$ supported in the daughter-energy sector,
we define $\langle O\rangle=\operatorname{Tr}[\rho(L)O]$,
with $\rho(L)$ normalized per source neutrino.
A subscript $\zeta$ specifies the environmental overlap.

The Hermitian operators
\begin{align}
 X_{13}^{L}&=\ket{1,L}\!\bra{3,L}+\ket{3,L}\!\bra{1,L},\nonumber\\
 Y_{13}^{L}&=-\ii\ket{1,L}\!\bra{3,L}
             +\ii\ket{3,L}\!\bra{1,L}
 \label{eq:quadratures}
\end{align}
give the two coherence quadratures, proportional to its real and
imaginary parts:
$\langle X_{13}^{L}\rangle=2\operatorname{Re}\rho_{13}^{L}$ and
$\langle Y_{13}^{L}\rangle=-2\operatorname{Im}\rho_{13}^{L}$.
Its magnitude is quantified by $\mathcal C_{13}=2|\rho_{13}^{L}|$ and its visibility by $\mathcal V_{13}=\mathcal C_{13}/(p_1+p_3)$.
The visibility is defined for nonzero daughter population.
The triangle inequality applied to eq.~\eqref{eq:integral} implies
$|F_{13}|\leq D$, so $0\leq\mathcal V_{13}\leq1$.
Even at $\zeta=1$, averaging over decay position reduces the
visibility when the daughter propagation phases vary appreciably.

The diagonal daughter contributions cancel in the neutrino--antineutrino
appearance difference, leaving the product of the imaginary mixing
factor and the imaginary propagated coherence.
With $c_e=U_{e1}U_{e3}^{*}$, we introduce the flavor-probability and coherence observables,
\begin{align}
 \mathcal W_P(\zeta)
 &=\ACP_{\mu e,\daughter}(\zeta)
   -\ACP_{\mu e,\daughter}(0),\nonumber\\
 \mathcal W_Y(\zeta)
 &=2\operatorname{Im}(c_e)\langle Y_{13}^{L}\rangle_\zeta.
 \label{eq:readouts}
\end{align}
Substitution into eq.~\eqref{eq:daughter-cp} establishes
\begin{equation}
 \mathcal W_P(\zeta)=\mathcal W_Y(\zeta)
 =-4\operatorname{Im}(c_e)\operatorname{Im}\rho_{13}^{L}(\zeta)
 =\zeta\,\ACP_{\mu e,\daughter}(\zeta=1).
 \label{eq:readout-identity}
\end{equation}
Positivity of the daughter block further bounds the observable:
\begin{equation}
 |\mathcal W_P|\leq
 2|\operatorname{Im}c_e|\mathcal C_{13}
 \leq4|\operatorname{Im}c_e|\sqrt{p_1p_3}.
 \label{eq:coherence-bound}
\end{equation}
This bound separates the available coherence from the quadrature
selected by CP. Coherence alone fixes neither the CP-odd mixing factor
nor its propagation quadrature, as the $4\to1,2$ null channel illustrates.

\section{Benchmark parameters and quantum circuits}
\label{sec:quantum}

We use the effective two-energy benchmark throughout the circuit study.
Table~\ref{tab:benchmark} lists the parameter values used in the calculation.
The active mixing angles, solar and atmospheric mass-squared differences,
and $\delta_{13}=-0.8\pi$ follow the normal-ordering reference values
used in ref.~\cite{Fiza:2021gvq}. We also adopt its smaller sterile-angle
set and $\Delta m_{41}^2=1.3\,\mathrm{eV}^2$, with the rotation order
in eq.~\eqref{eq:mixing}. These numerical inputs connect the calculation
to an established phenomenological benchmark.
We take $\delta_{14}=\delta_{24}=0$ to isolate the dependence on
$\delta_{13}$. The value of $\theta_{34}$ is retained from the reference
parameter set; it does not enter the vacuum $\nu_\mu\to\nu_e$
probabilities in this convention.

We choose $E_H=1\,\mathrm{GeV}$ as a convenient reference energy and
represent the daughter sector by $E_L=E_H/2=0.5\,\mathrm{GeV}$.
The ratio $E_L/E_H=1/2$ provides a simple benchmark for energy loss,
while the two energy sectors can be encoded in a single qubit.
A decay exponent $\kappa=1$ leaves both sectors
populated: an initial $\nu_4$ survives with probability $e^{-1}$ and
decays with probability $D\simeq0.632$.
We choose $\phi_{31}=\pi/2$ to retain a substantial post-decay phase.
Equal branching fractions maximize the interference factor
$\sqrt{B_1B_3}$ at fixed total width.

We evaluate the regenerated CP asymmetry at five equally spaced values
of $\zeta$ to test its predicted linear dependence on the environmental
overlap. The processor measurements compare $\zeta=0$ and $1$, which
maximize the difference in the regenerated contribution while preserving
the daughter populations. For the pair-$(1,2)$ control, equal branching
fractions permit daughter coherence, while the real mixing factor
ensures a vanishing regenerated CP asymmetry. A second control sets
$\delta_{13}=0$, so that all mixing phases vanish. Both controls retain
the benchmark baseline, energies, and total decay width.

\begin{table}[tbp]
\centering\small
\begin{tabular}{@{}p{0.31\textwidth}p{0.64\textwidth}@{}}
\toprule
Quantity & Benchmark values \\
\midrule
Active mixing angles &
$(\theta_{12},\theta_{13},\theta_{23})
=(34.3^\circ,8.58^\circ,48.8^\circ)$ \\
Sterile mixing angles &
$(\theta_{14},\theta_{24},\theta_{34})
=(5.7^\circ,5.0^\circ,20.0^\circ)$ \\
Mass-squared differences &
$(\Delta m_{21}^2,\Delta m_{31}^2,\Delta m_{41}^2)
=(7.5\times10^{-5},2.56\times10^{-3},1.3)\,\mathrm{eV}^2$ \\
Mixing phases &
$(\delta_{13},\delta_{14},\delta_{24})=(-0.8\pi,0,0)$ \\
Energies and propagation &
$(E_H,E_L)=(1,0.5)\,\mathrm{GeV}$;
$\kappa=1$; $\phi_{31}=\pi/2$ \\
Baseline and decay parameter &
$L\simeq121.078\,\mathrm{km}$;
$\alpha_4\simeq1.630\times10^{-3}\,\mathrm{eV}^2$ \\
Branching fractions &
$B_1=B_3=1/2$; pair control: $B_1=B_2=1/2$ \\
Environmental overlap &
Simulation: $0,0.25,0.5,0.75,1$; processor: $0,1$ \\
\bottomrule
\end{tabular}
\caption{Parameters of the effective two-energy model used to compare
classical predictions with quantum measurements. The mixing angles,
mass-squared differences, and $\delta_{13}$ follow the normal-ordering
inputs and smaller sterile-angle set of ref.~\cite{Fiza:2021gvq}.
The chosen energies and decay parameters retain appreciable parent
survival and daughter interference, while variation of the environmental
overlap isolates the coherence dependence.
The baseline $L$ and decay parameter $\alpha_4$ follow from
$\phi_{31}$ and $\kappa$ through eq.~\eqref{eq:units}.}
\label{tab:benchmark}
\end{table}

The splitting $\Delta m_{31}^2$ sets the daughter oscillation scale, while $\Delta m_{41}^2$ governs the rapidly varying parent interference. The continuous-spectrum calculation in section~\ref{sec:results}
extends this benchmark by replacing the single daughter energy with an energy distribution.

The complete two-energy channel maps states in the four-dimensional
source mass space to the combined parent and daughter energy sectors.
It is implemented by a unitary circuit
on two mass qubits, one energy qubit, and two environment qubits;
tracing out the environment recovers the decay channel. This
Stinespring dilation~\cite{Stinespring:1955eig,Iten:2016uzx} is
constructed in appendix~\ref{app:circuit} and checked against
independent evolution of eq.~\eqref{eq:master} in
appendix~\ref{app:validation}.

The regenerated contribution depends on the source only through
$p_{\mu4}^{\mathrm{in}}$. We divide the unnormalized daughter density matrix by its trace,
$w_D=p_{\mu4}^{\mathrm{in}}D$, to obtain the state conditioned on decay. For equal branching fractions,
\begin{equation}
 \rho_D(\zeta)=\frac{\rho_{\mu,D}}{w_D}
 =\frac12\begin{pmatrix}
 1&r_\zeta e^{\ii\chi}\\ r_\zeta e^{-\ii\chi}&1
 \end{pmatrix},\qquad
 r_\zeta=\frac{\zeta|F_{13}|}{D},\qquad \chi=\arg F_{13}.
 \label{eq:conditional}
\end{equation}
Here $\rho_{\mu,D}$ is the unnormalized daughter block, the basis is
$(\ket{\nu_1},\ket{\nu_3})$, and $r_\zeta=\mathcal V_{13}$.
The eigenvalues are $(1\pm r_\zeta)/2$. At $\zeta=1$, decay-position
averaging gives $r_1\simeq0.905$, so the smaller eigenvalue remains
approximately $0.047$ and the daughter state is mixed.

\begin{figure}[tbp]
\centering
\includegraphics[page=1,width=0.98\textwidth]{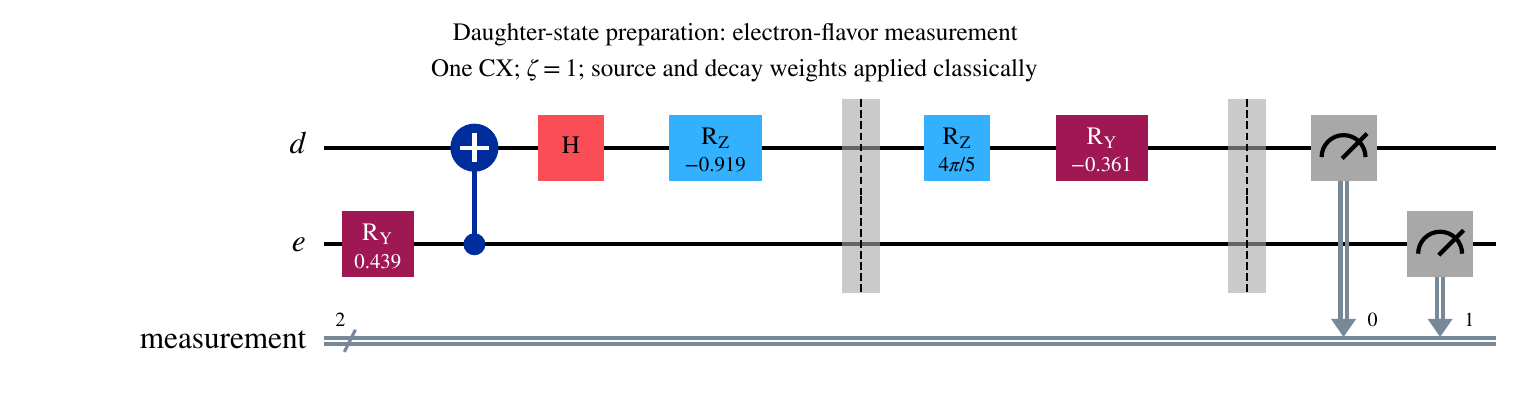}\\[0.7em]
\includegraphics[page=2,width=0.98\textwidth]{logical_circuits.pdf}
\caption{Compact circuits at $\zeta=1$ for electron-flavor projection (upper) and the $Y$ quadrature (lower). The qubit $d$ encodes the ordered daughter pair, while the auxiliary
qubit $e$ completes its purification. The preparation angles are $\beta_1\simeq0.439$ and $\chi\simeq0.919$, in radians. The flavor rotation uses the normalized projection in eq.~\eqref{eq:conditional-readout};
$S^\dagger=\diag(1,-\ii)$ followed by the Hadamard gate $H$ selects the $Y$ basis. Environment outcomes are summed.}
\label{fig:compact-circuit}
\end{figure}

We encode the daughter pair in qubit $d$ and prepare a pure joint state
with an environment qubit $e$, whose partial trace gives $\rho_D$.
Starting from $\ket{0}_e\ket{0}_d$, this purification is prepared by
$R_y(\beta_\zeta)$ on $e$, a controlled-NOT (CNOT) gate
$\CX_{e\to d}$, and then a Hadamard gate and $R_z(-\chi)$ on $d$, with
\begin{equation}
 \beta_\zeta=2\arcsin\sqrt{\frac{1-r_\zeta}{2}},\qquad
 w_D\simeq4.754\times10^{-3}.
 \label{eq:compact-angle}
\end{equation}
The single-qubit rotations obey
$R_a(\vartheta)=\exp(-\ii\vartheta\sigma_a/2)$ for the Pauli matrix
$\sigma_a$ along $a=y,z$. Each shot is one circuit execution followed
by measurement. We sum over the environment outcomes, so every shot
contributes. Figure~\ref{fig:compact-circuit} shows the two measurement
bases.

For a daughter pair $(j,k)$, we define
$n_e^{jk}=|U_{ej}|^2+|U_{ek}|^2$ and the normalized electron projection
$\ket{e_D^\nu}=(U_{ej}^*\ket0+U_{ek}^*\ket1)/\sqrt{n_e^{jk}}$.
The antineutrino projection contains the conjugated coefficients.
The daughter appearance probabilities and the coherence-based CP
observable are then
\begin{equation}
 P_{\mu e,D}^{\nu,\bar\nu}
 =w_D n_e^{jk}\bra{e_D^{\nu,\bar\nu}}\rho_D
                 \ket{e_D^{\nu,\bar\nu}},\qquad
 \mathcal W_Y=2w_D\operatorname{Im}(c_e)\Tr(\rho_D\sigma_y).
 \label{eq:conditional-readout}
\end{equation}
The second expression applies to $(j,k)=(1,3)$.
We determine $w_D$ and $n_e^{jk}$ from the reference benchmark before
running the circuits and use them to convert the measured quantities
to the physical normalization.

For the reference benchmark,
$\ACP_{\mu e,\daughter}(\zeta=1)/w_D\simeq0.102$.
We prepare the conditional daughter state directly so that all
measurement shots probe the regenerated sector. The compiled circuit
contains one controlled-$Z$ gate,
$\mathrm{CZ}=\diag(1,1,1,-1)$, supported directly by the processor.

\section{Numerical results and quantum-computer measurements}
\label{sec:results}

\subsection{Continuous daughter spectrum}

We first examine the continuous daughter-energy spectrum,
keeping the source energy, baseline, total decay width,
mixing parameters, and branching fractions fixed at their
reference-benchmark values.
We take the lightest neutrino mass to be $m_1=0.01\,\mathrm{eV}$.
Together with the mass-squared differences, it determines the
daughter-energy endpoints specified in appendix~\ref{app:conventions}.
Figure~\ref{fig:spectrum} shows the regenerated
electron-appearance spectra for neutrinos and antineutrinos.
Their difference arises from the CP-sensitive flavor
projection of the propagated daughter coherence.

\begin{figure}[htbp]
\centering
\includegraphics[width=\textwidth]{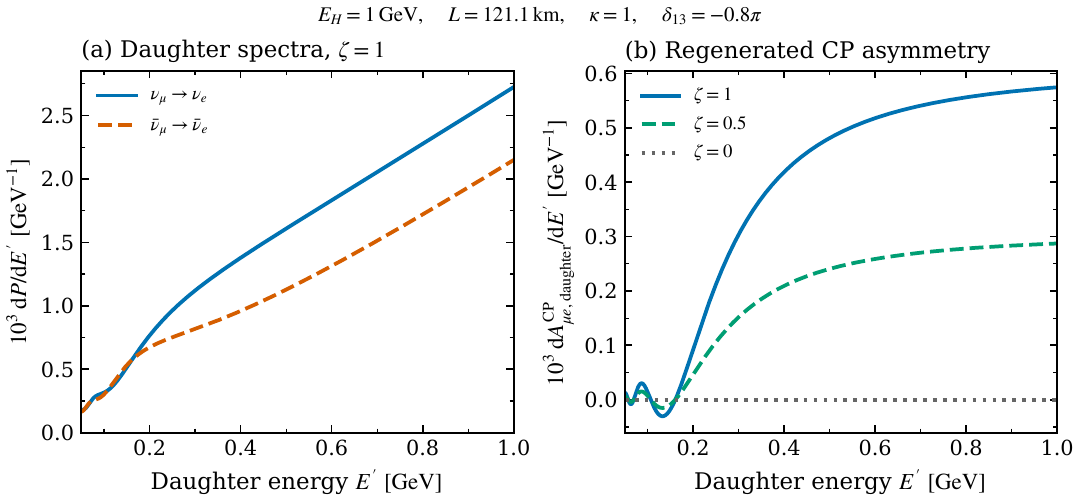}
\caption{
Continuous daughter spectra for the reference source and
propagation parameters.
Panel (a) shows neutrino and antineutrino electron-appearance
densities at $\zeta=1$; panel (b) shows their difference at
$\zeta=1,0.5,0$.
The spectral profiles follow eq.~\eqref{eq:spectral-profile},
with $m_1=0.01\,\mathrm{eV}$ and equal branching fractions
into $\nu_1$ and $\nu_3$.
The plotted interval is $0.05\leq E'/\mathrm{GeV}\leq1$,
and vertical-axis values scaled by $10^3$.
Energy-bin convergence is checked in appendix~\ref{app:conventions}.
}
\label{fig:spectrum}
\end{figure}

At lower daughter energies, the relative propagation phase
$\phi_{31}(E')=\Delta m_{31}^2L/(2E')$ increases.
The imaginary part of the interference kernel changes sign,
producing the low-energy structure in
figure~\ref{fig:spectrum}(b).
Averaging over the decay position suppresses the rapidly
oscillating contribution.
At fixed energy, reducing the environmental overlap $\zeta$
scales the regenerated CP difference without changing the
daughter mass populations.
The integrated asymmetry over the plotted interval is
\begin{equation}
 \int_{0.05\,\mathrm{GeV}}^{1\,\mathrm{GeV}}
 \frac{\dd\ACP_{\mu e,\daughter}}{\dd E'}\,\dd E'
 =
 3.753\times10^{-4}
 \qquad(\zeta=1).
 \label{eq:continuous-result}
\end{equation}
This energy-integrated result differs from the two-energy benchmark,
which places the regenerated population at $E_L$ and samples a single
daughter-energy phase.

\subsection{Two-energy benchmark}

For this benchmark, the initial unstable-state population is
$p_{\mu4}^{\mathrm{in}}\simeq7.521\times10^{-3}$.
At maximal environmental overlap, the surviving population,
regenerated populations, coherence, and visibility are
\begin{equation}
 \begin{aligned}
 p_4&\simeq2.767\times10^{-3},
 &p_1=p_3&\simeq2.377\times10^{-3},\\
 \mathcal C_{13}&\simeq4.304\times10^{-3},
 &\mathcal V_{13}&\simeq0.905.
 \end{aligned}
 \label{eq:b2-populations}
\end{equation}
The pair-$(1,2)$ control undergoes less decay-position averaging and
retains a larger coherence, $\mathcal C_{12}\simeq4.754\times10^{-3}$.
Its regenerated CP contribution nevertheless vanishes because
$\operatorname{Im}Z_{e;12}=0$ for the specified mixing matrix and real
decay amplitudes.

For the reference benchmark at $\zeta=1$, the regenerated CP asymmetry,
the parent contribution, and the total decay-induced shift are
\begin{equation}
 \begin{aligned}
 S_{\CP}\equiv\ACP_{\mu e,\daughter}(\zeta=1)
 &=4.855\times10^{-4},\\
 \DACP_{\mu e,\parent}
 &=5.692\times10^{-4},\\
 \DACP_{\mu e}(\zeta=1)
 &=1.055\times10^{-3}.
 \end{aligned}
 \label{eq:b2-total}
\end{equation}
Daughter regeneration accounts for approximately $46\%$ of
the total shift. Both contributions are positive at this benchmark, although their signs depend on different combinations of phases. Reducing $\zeta$ suppresses the regenerated term while leaving parent attenuation unchanged.
The ideal full-channel simulation in appendix~\ref{app:validation} reproduces this separation.
We use $S_{\CP}$ as the reference scale for the experimental
precision tests.

\subsection{Quantum-computer measurements}

We implemented the compact daughter-state circuits on IBM Kingston
at $\zeta=0$ and $1$. Table~\ref{tab:resources-main} summarizes the
experimental resources.

\begin{table}[tbp]
\centering
\small
\begin{tabular}{@{}ll@{}}
\toprule
Quantity & Implementation \\
\midrule
Quantum computer & IBM Kingston \\
Circuit & Two-qubit daughter-state purification \\
Entangling gates & One native CZ per circuit \\
Environmental overlap & $\zeta=0,1$ \\
Observable measurements & $327\,680$ shots across 16 circuit settings \\
Calibration measurements & $32\,768$ shots across two runs \\
Measurement correction & Inversion of the two-qubit calibration matrix \\
Uncertainty & Measurement and calibration covariance \\
\bottomrule
\end{tabular}
\caption{
Resources for the endpoint experiment.
Four observable-measurement circuits receive $32\,768$
shots each and the remaining twelve receive $16\,384$ each.
Each computational-basis calibration state receives
$4\,096$ shots before and after the observable measurements.
The environment outcomes are summed in every estimator.
}
\label{tab:resources-main}
\end{table}

We correct measurement errors using computational-basis
calibration~\cite{Maciejewski:2019jor,Bravyi:2020bpr}.
The error bars show one standard error, accounting for statistical uncertainties from measurement and calibration and retaining correlations from shared data.
The calibration and repeated measurement blocks satisfy the predefined
stability criteria. Appendix~\ref{app:calibration} describes measurement calibration and error mitigation, while appendix~\ref{app:statistics} details the calculation of statistical uncertainties.

\begin{figure}[tbp]
\centering
\includegraphics[width=\textwidth]{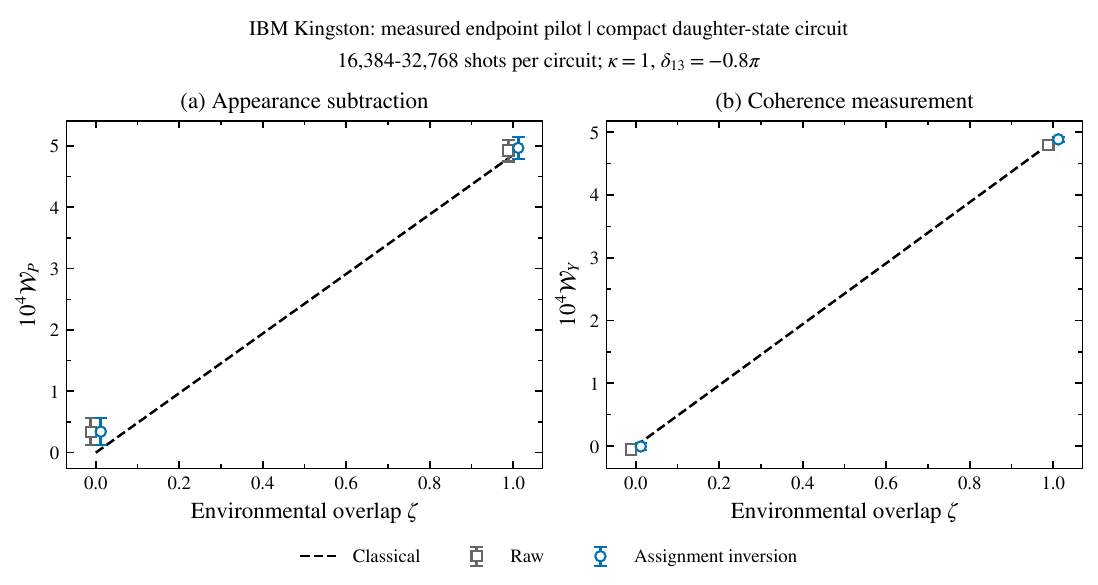}
\caption{
IBM Kingston measurements of $\mathcal W_P$ and
$\mathcal W_Y$ at $\zeta=0,1$. Grey squares denote raw estimates and blue circles denote corrected estimates. Error bars represent one standard error.
Dashed curves show the classical prediction. All estimates include the normalization $w_D$; the flavor-probability estimates also include $n_e^{13}$. Horizontal offsets separate overlapping symbols.
}
\label{fig:ibm-cp}
\end{figure}

We reconstruct the regenerated CP contribution from flavor probabilities
and the $Y$ expectation value using eq.~\eqref{eq:conditional-readout}.
Figure~\ref{fig:ibm-cp} compares both estimates with the theoretical
prediction, with points representing the measurements and curves
showing the predicted dependence.

\begin{table}[tbp]
\centering
\begin{tabular}{@{}lccc@{}}
\toprule
Estimator at $\zeta=1$ & Raw & Corrected & Residual$/\sigma$ \\
\midrule
$\mathcal W_P$
& $4.923\pm0.177$ & $4.967\pm0.179$ & $+0.63$ \\
$\mathcal W_Y$
& $4.795\pm0.037$ & $4.884\pm0.039$ & $+0.74$ \\
$\ACP_{\mu e,\daughter}$
& $4.934\pm0.124$ & $4.979\pm0.125$ & $+0.99$ \\
\bottomrule
\end{tabular}
\caption{
Coherent-endpoint estimates in units of $10^{-4}$,
with one-standard-error uncertainties.
The classical prediction is $4.855$ in these units.
The final column reports the corrected residual divided
by its standard error $\sigma$.
The last row is the direct appearance difference without
incoherent-reference subtraction and shares counts with
$\mathcal W_P$.
}
\label{tab:ibm-results}
\end{table}

Both primary estimates agree with the prediction within
one standard error.
Their difference,
$(0.083\pm0.183)\times10^{-4}$,
is $0.45$ standard errors after covariance propagation.
The coherence measurement reaches a statistical uncertainty
of $0.80\%$ of $S_{\CP}$.
It determines the relevant interference component directly,
whereas the flavor-probability estimator combines four
sampled probabilities whose counting variances accumulate.
The direct appearance difference in
table~\ref{tab:ibm-results} also agrees with the prediction
before incoherent-reference subtraction.

\begin{figure}[tbp]
\centering
\includegraphics[width=\textwidth]{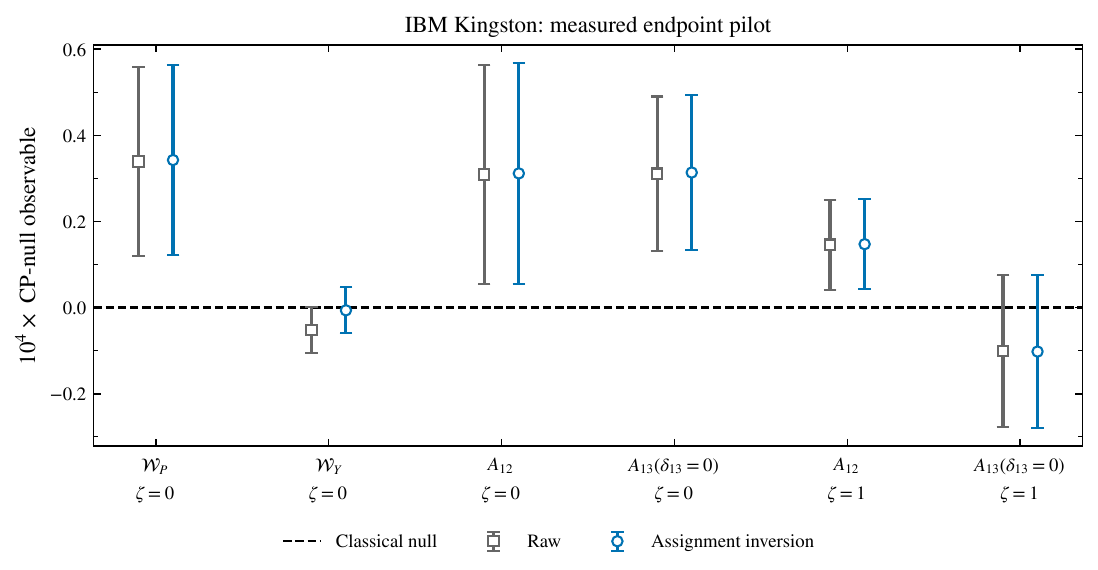}
\caption{
Six predefined null controls:
$\mathcal W_P$ and $\mathcal W_Y$ at $\zeta=0$,
and the pair-$(1,2)$ and $\delta_{13}=0$ appearance
differences at both endpoints.
Symbols distinguish raw and corrected estimates.
Error bars represent one standard error; the dashed line
denotes zero.
The predefined precision criterion is
$|\widehat A|+1.96\sigma_A<0.20S_{\CP}$.
Individual values appear in table~\ref{tab:nulls}.
}
\label{fig:ibm-nulls}
\end{figure}

The null controls separate the physical requirements for
a regenerated CP signal.
At $\zeta=0$, the daughter coherence vanishes.
The pair-$(1,2)$ control retains coherence but has no imaginary
electron-flavor mixing factor.
Setting $\delta_{13}=0$ removes that factor from the
pair-$(1,3)$ contribution.
Figure~\ref{fig:ibm-nulls} compares these configurations with
their zero predictions.

All six controls are consistent with zero within two standard
errors and satisfy the predefined precision criterion.
Their absolute central values plus $1.96$ standard errors
range from $2.3\%$ to $16.8\%$ of $S_{\CP}$.
The measurements thus resolve the suppression of the
regenerated contribution when either daughter coherence
or its CP-sensitive mixing factor vanishes. Appendix~\ref{app:statistics} also reports the direct appearance difference at $\zeta=0$ as an additional consistency check.

\section{Discussion and conclusions}
\label{sec:discussion}
\label{sec:conclusions}

The central physics question was how a CP-symmetric decay interaction changes an asymmetry generated by neutrino mixing. Parent attenuation changes the interference involving the unstable state, while visible regeneration introduces daughter interference at lower energy. Separating these contributions identifies the origin of the
decay-induced shift. The coherent pair $(\nu_1,\nu_2)$ provides a particularly useful comparison: its regenerated CP contribution vanishes for the specified mixing matrix and real decay amplitudes, despite its nonzero coherence. The regenerated CP asymmetry therefore depends on daughter coherence, a CP-odd mixing factor, and the relative propagation phase accumulated after decay.

For the reference benchmark, daughter regeneration contributes $4.85\times10^{-4}$, approximately $46\%$ of the total decay-induced CP shift. On the quantum processor, flavor-probability and coherence measurements give $(4.97\pm0.18)\times10^{-4}$ and
$(4.884\pm0.039)\times10^{-4}$, respectively. The uncertainties are one standard error and include observable and calibration statistics. Both measurements agree with the prediction, while the six predefined null controls confirm the suppression of the regenerated signal within
their statistical precision.

The environmental overlap offers a direct connection to the
microscopic physics of decay. Retaining the scalar recoil and the neutrino wave packets before tracing out the unobserved degrees of freedom would relate this overlap to the daughter energies, mass splittings, and localization scales \cite{Akhmedov:2009rb,Akhmedov:2010ua,Giunti:1997wq,Visinelli:2008ds,Farzan:2008eg,Akhmedov:2022bjs}.
Such a calculation could establish where daughter coherence
survives under specific production and detection conditions.

The connection to neutrino experiments also depends on how
regeneration changes the observed energy spectrum.
To predict event spectra, we would extend the propagation model to
include matter effects and combine the resulting probabilities with
the incident flux, interaction cross sections, and detector response
\cite{Gago:2017zzy,Coloma:2017zpg,Porto-Silva:2020gma}.
A joint analysis of neutrino and antineutrino spectra could then
explore how intrinsic CP phases and decay affect the signal in the
presence of the neutrino--antineutrino asymmetry induced by ordinary
matter \cite{PhysRevD.17.2369}.
Daughter events at lower reconstructed energies may change the inferred
dependence on the CP phases even when parent disappearance is well
constrained.

A complete-channel experiment on a quantum computer would extend the present conditional-state measurement.
The five-qubit construction in appendix~\ref{app:circuit} provides a starting point. An unstable-state input could first test survival and energy transfer, followed by coherent mass-state inputs that test parent interference. Preparing the physical muon-flavor state would then bring both contributions into the same experiment. The environment would remain unobserved throughout the analysis. The small daughter population makes gate accuracy and shot allocation
central to this extension. Measurements at intermediate values of the environmental overlap
would test the predicted linear dependence of the regenerated
CP asymmetry on daughter coherence.

Visible decay could also be incorporated into interacting
many-neutrino systems. In that case quantum simulation may offer a computational advantage in regimes where the resulting correlations make accurate classical evolution prohibitively expensive. The decay-channel construction developed here could form one component of such a many-body simulation.

We hope that this work will support further studies of the interplay
between visible decay, coherence, and CP violation in neutrino
propagation. As neutrino experiments and quantum processors advance,
controlled simulations may offer complementary insight into the
quantum dynamics underlying neutrino signals.

\begin{acknowledgments}
AS and DS are partially supported by the U.S. National Science Foundation (NSF) under Grant No.~PHY-2310363. AS also acknowledges support under NSF Grant No.~OAC-2417682 and from a Universities Research Association Visiting Scholars Fellowship at Fermilab. AS thanks Joachim Kopp for introducing him to this topic, for useful discussions during a visit to CERN, and for valuable comments on the work. We acknowledge the use of ChatGPT (OpenAI) for assistance with coding and sentence-level editing.
\end{acknowledgments}

\appendix

\section{Mixing convention and continuous spectral input}
\label{app:conventions}

Each rotation in eq.~\eqref{eq:mixing} acts as the identity outside
the $i$--$j$ plane. Its nontrivial block is
\begin{equation}
 \left.\widetilde R_{ij}(\theta_{ij},\delta_{ij})\right|_{ij}
 =\begin{pmatrix}
 \cos\theta_{ij}&\sin\theta_{ij}e^{-\ii\delta_{ij}}\\
 -\sin\theta_{ij}e^{\ii\delta_{ij}}&\cos\theta_{ij}
 \end{pmatrix},\qquad
 R_{ij}=\widetilde R_{ij}(\theta_{ij},0).
 \label{eq:rotation}
\end{equation}
With $c_{ij}=\cos\theta_{ij}$ and $s_{ij}=\sin\theta_{ij}$, the
entries relevant to the regenerated $\mu e$ signal are
\begin{equation}
 \begin{aligned}
 U_{e1}&=c_{14}c_{13}c_{12},& U_{e2}&=c_{14}c_{13}s_{12},\\
 U_{e3}&=c_{14}s_{13}e^{-\ii\delta_{13}},&
 U_{\mu4}&=c_{14}s_{24}e^{-\ii\delta_{24}}.
 \end{aligned}
 \label{eq:mixing-entries}
\end{equation}
The fourth electron entry is $U_{e4}=s_{14}e^{-\ii\delta_{14}}$.
It enters the parent quartets in eq.~\eqref{eq:parent-shift}.
The daughter mixing factor is
$\operatorname{Im}c_e=c_{14}^2c_{13}c_{12}s_{13}\sin\delta_{13}$.

A mass-basis rephasing $\ket{\nu_i}\to e^{\ii\xi_i}\ket{\nu_i}$,
with real phases $\xi_i$, transforms the parameters as
\begin{equation}
 U_{\alpha i}\to e^{\ii\xi_i}U_{\alpha i},\qquad
 \gamma_{4j}^{(nm)}\to
 e^{\ii(\xi_4-\xi_j)}\gamma_{4j}^{(nm)}.
 \label{eq:rephasing}
\end{equation}
The phases cancel in $Z_{\beta;jk}^{(nm)}$.
The decay sector is specified by real jump amplitudes after the mixing convention has been fixed.
The zero regenerated asymmetry for $4\to1,2$ refers to this joint choice of mixing and decay parameters.

In practical units, the dimensionless propagation phases and decay
exponents are
\begin{equation}
 \Phi_{in}\simeq2.534\,
 \frac{\Delta m_{i1}^2}{\mathrm{eV}^2}
 \frac{L/\mathrm{km}}{E_n/\mathrm{GeV}},
 \qquad
 \kappa_n\simeq5.068\,
 \frac{\alpha_4}{\mathrm{eV}^2}
 \frac{L/\mathrm{km}}{E_n/\mathrm{GeV}}.
 \label{eq:units}
\end{equation}
The first coefficient multiplies the amplitude phase.
The argument of an oscillation term $\sin^2(\Delta m^2L/4E)$ contains half this coefficient.

For the continuous spectrum, we combine the lepton-number-conserving
scalar and pseudoscalar differential decay rates with equal coefficients
\cite{Lindner:2001fx,Gago:2017zzy,Kopp:2026tnx}.
In terms of the squared mass ratio $q_j=m_j^2/m_4^2$ and the
daughter-energy fraction $x=E'/E_H$, the normalized daughter spectrum is
\begin{equation}
 f_j(E')=
 \frac{x+q_j/x}
 {E_H\left[\frac12(1-q_j^2)-q_j\ln q_j\right]}
 \Theta(x-q_j)\Theta(1-x),\qquad
 \int f_j(E')\,\dd E'=1,
 \label{eq:spectral-profile}
\end{equation}
where $\Theta$ is the Heaviside function.
The masses obey $m_j^2=m_1^2+\Delta m_{j1}^2$.
We obtain the diagonal differential decay rates from the normalized
profile as $\eta_{4j}(E_H,E')=\Gamma_4(E_H)B_j f_j(E')$.
Their off-diagonal source is separately specified by the real overlap $\zeta$. This choice produces the differential daughter matrix
\begin{align}
 \frac{\dd\rho_{jj}^{\daughter}}{\dd E'}
 &=p_{\mu4}^{\mathrm{in}}B_jD f_j(E'),\nonumber\\
 \frac{\dd\rho_{13}^{\daughter}}{\dd E'}
 &=\zeta p_{\mu4}^{\mathrm{in}}
   \sqrt{B_1B_3 f_1(E')f_3(E')}\,
   F_{13}\bigl(\kappa,\phi_{31}(E')\bigr).
 \label{eq:spectral-density}
\end{align}
The product $f_1f_3$ restricts interference to daughter energies allowed
in both decay channels. The effective decay amplitudes in this construction obey the same CP-conjugation rule as eq.~\eqref{eq:daughter-cp}.

We integrate over $E'\in[0.05,1]\,\mathrm{GeV}$ for
figure~\ref{fig:spectrum}. Within the kinematically allowed energy range, the cumulative
distribution is
\begin{equation}
 F_j^{\mathrm{cum}}(E')=
 \frac{\frac12(x^2-q_j^2)+q_j\ln(x/q_j)}
 {\frac12(1-q_j^2)-q_j\ln q_j}.
 \label{eq:cumulative}
\end{equation}
It equals zero below $q_jE_H$ and one above $E_H$.
The omitted energy interval carries total neutrino probability
\begin{equation}
 P_{\mathrm{sink}}=
 p_{\mu4}^{\mathrm{in}}D\sum_{j=1,3}B_j
 F_j^{\mathrm{cum}}(0.05\,\mathrm{GeV})
 \simeq4.44\times10^{-5}.
 \label{eq:sink}
\end{equation}
Including this population, the sum over all four flavors and all
energy blocks is unity.

For the convergence check, we integrate each diagonal rate over
logarithmic energy bins and evaluate the propagation kernel at the
arithmetic bin centers. The cross term uses the geometric mean of
the two integrated partial rates, as in
eq.~\eqref{eq:overlap-jumps}. Direct numerical integration of
eq.~\eqref{eq:spectral-density} over daughter energy provides the
reference. With 128 bins, both integrated appearance probabilities
and their difference agree with it to within $7.65\times10^{-9}$.

\section{Finite-baseline channel and elementary circuit}
\label{app:circuit}

We encode mass in $(m_1,m_0)$ as
$\nu_1\leftrightarrow00$, $\nu_2\leftrightarrow01$,
$\nu_3\leftrightarrow10$, and $\nu_4\leftrightarrow11$.
The energy qubit $E$ stores $E_H$ in zero and $E_L$ in one.
Two additional qubits, $a$ and $b$, encode the environment records.
The energy qubit and both environment qubits are initialized in
$\ket{0}$.
The full map acts on an arbitrary mass density matrix $\rho_H$ at $E_H$,
\begin{equation}
 \mathcal E_{L,\zeta}(\rho_H)=\sum_{r=0,+,-}K_r\rho_HK_r^\dagger.
 \label{eq:full-channel}
\end{equation}
We write $\Phi_{iH}=\Delta m_{i1}^2L/(2E_H)$ for the parent-energy
propagation phases. The following Kraus operators contain both parent propagation and the integrated daughter evolution~\cite{Stinespring:1955eig,CHOI1975285}.

For equal branching fractions, the daughter block produced by a pure $\nu_4$ input is
\begin{equation}
 Q_\zeta=\frac12
 \begin{pmatrix}D&\zeta F_{13}\\\zeta F_{13}^{*}&D\end{pmatrix}
 \quad\hbox{in the basis }(\ket{1,L},\ket{3,L}).
 \label{eq:Q}
\end{equation}
Using $r_\zeta$ and $\chi$ from eq.~\eqref{eq:conditional}, the
normalized eigenvectors and weights are
\begin{equation}
 \ket{d_\pm}=\frac{\ket{\nu_1}\pm e^{-\ii\chi}\ket{\nu_3}}{\sqrt2},
 \qquad \lambda_\pm=\frac{1\pm r_\zeta}{2},
 \qquad
 Q_\zeta=D\sum_{\sigma=\pm}\lambda_\sigma
 \ket{d_\sigma,E_L}\!\bra{d_\sigma,E_L}.
 \label{eq:daughter-eigenvectors}
\end{equation}
The symbol $\sigma$ labels the two signs.
The phase $\chi$ originates entirely in the propagation integral. It is identical in the neutrino and antineutrino decay circuits.

A Kraus representation from the four-dimensional source mass space to the eight-dimensional mass--energy space is
\begin{align}
 K_0&=\sum_{i=1}^{3}e^{-\ii\Phi_{iH}}\ket{i,H}\!\bra{\nu_i}
   +e^{-\kappa/2-\ii\Phi_{4H}}\ket{4,H}\!\bra{\nu_4},\nonumber\\
 K_\pm&=\sqrt{D\lambda_\pm}\,e^{-\ii\Phi_{4H}}
            \ket{d_\pm,E_L}\!\bra{\nu_4},\qquad
 K_0^\dagger K_0+K_+^\dagger K_++K_-^\dagger K_-=\mathbb 1_4.
 \label{eq:kraus}
\end{align}
The common phase in $K_\pm$ matches the preparation convention and
cancels from the reduced map. At this benchmark,
$0\leq r_\zeta\leq r_1<1$, so both daughter weights are positive.
The three Kraus operators are linearly independent; the channel
therefore has Choi rank three, the minimum number of Kraus operators
required to represent it~\cite{CHOI1975285}.

The same rank follows from the general construction
$C_L=(\mathcal E_{L,\zeta}\otimes\mathbb 1)
\ket{\Omega}\!\bra{\Omega}$, where
$\ket{\Omega}=\sum_{i=1}^{4}\ket{\nu_i}\ket{\nu_i}$ is unnormalized. The identity acts on a copy of the source space, and $C_L$ is the Choi matrix on output times input space. The Kraus operators are recovered from its nonzero eigenvectors with the reshaping convention $\ket{K}\!\rangle=\sum_{a,i}K_{ai}\ket{a}_{\mathrm{out}}\ket{i}_{\mathrm{in}}$~\cite{CHOI1975285,Kopp:2026tnx,Havel:2002ceg,JAMIOLKOWSKI1972275}. Here $a$ labels an output mass--energy basis state.

\label{app:gates}

\begin{figure}[tbp]
 \centering
 \includegraphics[width=0.98\textwidth]{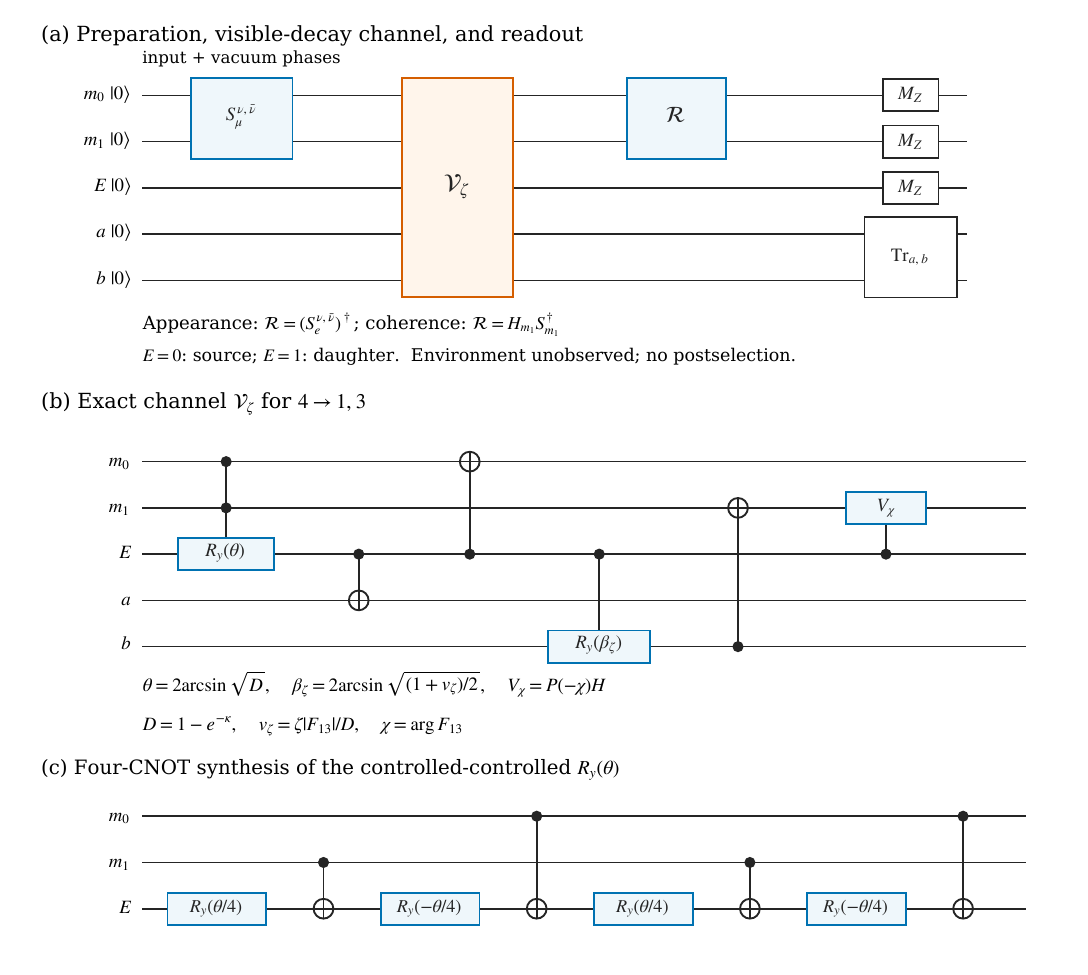}
 \caption{Five-qubit circuit for the $4\to1,3$ decay channel.
 (a) $S_\mu^{\nu,\bar\nu}$ prepares the source including the phases $\Phi_{iH}$, and $\mathcal R$ selects the measurement basis for appearance probabilities or coherence.
 The symbols $M_Z$ denote computational-basis measurements.
 (b) $V_\zeta$ acts on $(m_0,m_1,E,a,b)$ with $Eab$ initially zero. The drawing labels the full-channel eigenmode angle by $\beta_\zeta$; here it is written $\widetilde\beta_\zeta$ to distinguish it from
 the compact angle. The three angles are defined in
 eq.~\eqref{eq:gate-angles}; the block $V_\chi=P(-\chi)H$ applies the Hadamard gate $H$ followed by $P(-\chi)$.
 (c) Four controlled-NOT gates implement the controlled-controlled rotation exactly. We sum over all environment measurement outcomes.
}
 \label{fig:circuit-main}
\end{figure}

The circuit uses the register order
$(q_0,q_1,q_2,q_3,q_4)=(m_0,m_1,E,a,b)$.
Let $R_y(\vartheta)=\exp(-\ii\vartheta\sigma_y/2)$, where
$\sigma_y$ is the Pauli matrix, and let
$P(\varphi)=\diag(1,e^{\ii\varphi})$.
The Hadamard gate is
$H=2^{-1/2}\left(\begin{smallmatrix}1&1\\1&-1\end{smallmatrix}\right)$;
in circuit expressions this symbol denotes a gate rather than the
propagation Hamiltonian in eq.~\eqref{eq:hamiltonian}.
The three angles of the decay block are
\begin{equation}
 \theta=2\arcsin\sqrt D,\qquad
 \widetilde\beta_\zeta=2\arcsin\sqrt{\frac{1+r_\zeta}{2}},\qquad
 \chi=\arg F_{13}.
 \label{eq:gate-angles}
\end{equation}
Here, $\theta\simeq1.838$, $\chi\simeq0.919$, and
$\widetilde\beta_1\simeq2.703$, in radians. The full-channel construction orders the environment
eigenmodes oppositely to the compact purification, so
$\widetilde\beta_\zeta=\pi-\beta_\zeta$.

We apply $R_y(\theta)$ to the energy qubit $E$, conditioned on both
mass qubits being in $\ket{1}$. This rotation transfers the unstable-state
amplitude into the daughter energy level. A controlled-NOT gate from $E$ to $a$ records the decay.
A second controlled-NOT gate from $E$ to $m_0$ changes $m_0$ to
$\ket{0}$ in the decay branch, as required by the encoding of both
daughter states. The sequence then
applies controlled-$R_y(\widetilde\beta_\zeta)$ from $E$ to $b$,
$\CX_{b\to m_1}$, controlled-$H$ from $E$ to $m_1$, and
controlled-$P(-\chi)$ on the same pair.
For an unstable input the resulting isometry is
\begin{align}
 V_\zeta\ket{\nu_4}_m\ket{0}_E\ket{0}_a\ket{0}_b
 ={}&e^{-\kappa/2}\ket{\nu_4,E_H}\ket0_a\ket0_b\nonumber\\*
 &+\sqrt D\left(
 \sqrt{\lambda_-}\ket{d_-,E_L}\ket1_a\ket0_b
 +\sqrt{\lambda_+}\ket{d_+,E_L}\ket1_a\ket1_b\right).
 \label{eq:explicit-isometry}
\end{align}
In the input ket, $m$ denotes the two-qubit mass register.
The three light input masses remain at $E_H$ with the environment in
zero. Tracing $a,b$ reproduces eqs.~\eqref{eq:Q} and \eqref{eq:kraus}
after the preparation phases are included.
For the $4\to1,2$ control, we exchange the roles of $m_0$ and $m_1$
and replace $\phi_{31}$ by $\phi_{21}$ in the kernel.

The four-CNOT decomposition in figure~\ref{fig:circuit-main}(c)
applies rotations with successive angles
$\theta/4,-\theta/4,\theta/4,-\theta/4$ to $E$.
Interleaved CNOT controls follow the order $m_1,m_0,m_1,m_0$.
The sign of each $R_y$ rotation is reversed by conjugation with the target bit flip. Their angles cancel for mass inputs $00$, $01$, and
$10$, and sum to $\theta$ for $11$.
This is a uniformly controlled rotation
\cite{Mottonen:2004dis}.

The source rotations prepare
\begin{equation}
 \begin{aligned}
 S_\mu^\nu\ket{00}&=\sum_iU_{\mu i}^*e^{-\ii\Phi_{iH}}\ket{\nu_i},\\
 S_\mu^{\bar\nu}\ket{00}&=\sum_iU_{\mu i}e^{-\ii\Phi_{iH}}\ket{\nu_i}.
 \end{aligned}
 \label{eq:source-preparation}
\end{equation}
The same decay circuit follows in both cases.
To measure the appearance probability, we define
$S_e^\nu\ket{00}=\sum_iU_{ei}^{*}\ket{\nu_i}$ and
$S_e^{\bar\nu}\ket{00}=\sum_iU_{ei}\ket{\nu_i}$.
Applying the inverse preparation maps the electron-flavor projection
onto the computational-basis outcome $00$ in each energy sector.
For $Y_{13}^{L}$, we apply $S^\dagger=\diag(1,-\ii)$ to $m_1$
and then $H$. The bit strings $100$ and $110$ receive weights $+1$
and $-1$, respectively; all other strings receive zero.
Applying only $H$ measures $X_{13}^{L}$ with the same weights.
These weighted outcomes give the unnormalized daughter coherence
quadratures.
Here measured bit strings follow the order $Em_1m_0$.
The daughter electron probability is the frequency of $100$ after its
flavor rotation, with all shots in the normalization.

\begin{figure}[htbp]
 \centering
 \includegraphics[width=\textwidth]{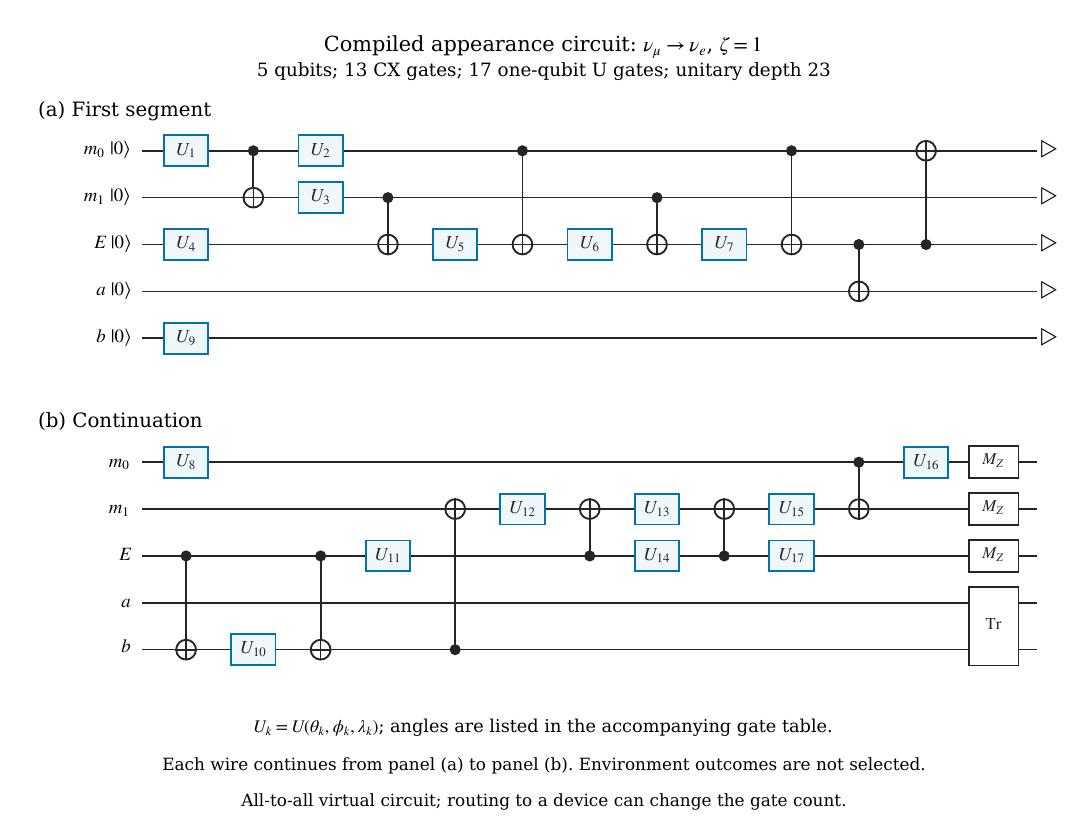}
 \caption{Compiled neutrino appearance circuit at $\zeta=1$.
The wires continue from panel (a) to panel (b).
Qiskit 2.5.2~\cite{Javadi-Abhari:2024kbf} compiles the source preparation, decay block, and electron-flavor projection into $13$ CNOTs and $17$ one-qubit gates at optimization level $3$, with exact synthesis and seed $20260919$.
The unitary depth is $23$, excluding terminal measurements.These resources refer to all-to-all logical connectivity.}
 \label{fig:circuit-compiled}
\end{figure}

The full-channel simulations use the structured decay circuit. Including source-state preparation and electron-flavor measurement increases the CNOT count from 11 to 13. We compile the same isometry with a general synthesis method to compare resource costs. The IBM measurements use the separate two-qubit preparation of the conditional daughter state, with one native CZ gate per circuit.

\begin{table}[htbp]
 \centering
 \begin{tabular}{@{}lrrr@{}}
 \toprule
 Circuit at $\zeta=1$ & CNOTs & One-qubit $U$ gates & Depth \\
 \midrule
 Structured decay block & 11 & 12 & 19 \\
 General synthesis of the same isometry & 32 & 39 & 65 \\
 Complete appearance circuit & 13 & 17 & 23 \\
 \bottomrule
 \end{tabular}
 \caption{Logical resources after exact compilation into
 $\{U,\CX\}$. The general-isometry comparison uses the same
 four-dimensional input space and five-qubit output as the structured decay block. State preparation and detection are included only in the final row. Measurements are excluded from the depths.}
 \label{tab:resources}
\end{table}

Table~\ref{tab:resources} compares the structured block with Qiskit's general isometry synthesis~\cite{Iten:2016uzx,Javadi-Abhari:2024kbf}. The daughter eigenbasis and the single unstable input reduce the two-qubit gate count from 32 to 11 for the decay block.
For source preparation, a singular-value decomposition of the
$2\times2$ amplitude matrix gives the two Schmidt coefficients and
the local basis rotations. A single $R_y$ rotation followed by a CNOT
prepares the entangled state with these coefficients; the local
rotations then place it in the required mass basis. Reversing this
construction implements the electron-flavor projection.

The one-qubit gate convention in figure~\ref{fig:circuit-compiled} is
\begin{equation}
 U(\vartheta,\varphi,\lambda)=
 \begin{pmatrix}
 \cos(\vartheta/2)&-e^{\ii\lambda}\sin(\vartheta/2)\\
 e^{\ii\varphi}\sin(\vartheta/2)&
 e^{\ii(\varphi+\lambda)}\cos(\vartheta/2)
 \end{pmatrix}.
 \label{eq:Ugate}
\end{equation}
The angles $\vartheta$, $\varphi$, and $\lambda$ specify a general
one-qubit unitary up to a global phase.
The ancillary gate-angle table and circuit files specify the compiled
gates; the circuit files retain the full numerical precision.

\FloatBarrier
\section{Native circuits and classical validation}
\label{app:validation}

With $\lambda_\pm=(1\pm r_\zeta)/2$, the state after the first
rotation and CNOT is
$\sqrt{\lambda_+}\ket{0}_d\ket{0}_e+
\sqrt{\lambda_-}\ket{1}_d\ket{1}_e$.
The final Hadamard and phase rotation map the two system basis states
to $\ket{d_+}$ and $\ket{d_-}$, respectively.
Tracing over $e$ therefore reproduces eq.~\eqref{eq:conditional}.
The phase rotation differs from $P(-\chi)$ only by an irrelevant
global phase. The gate $S^\dagger=\diag(1,-\ii)$ followed by $H$
rotates the Pauli-$Y$ measurement into the computational basis.

\begin{figure}[tbp]
\centering
\includegraphics[page=1,width=\textwidth]{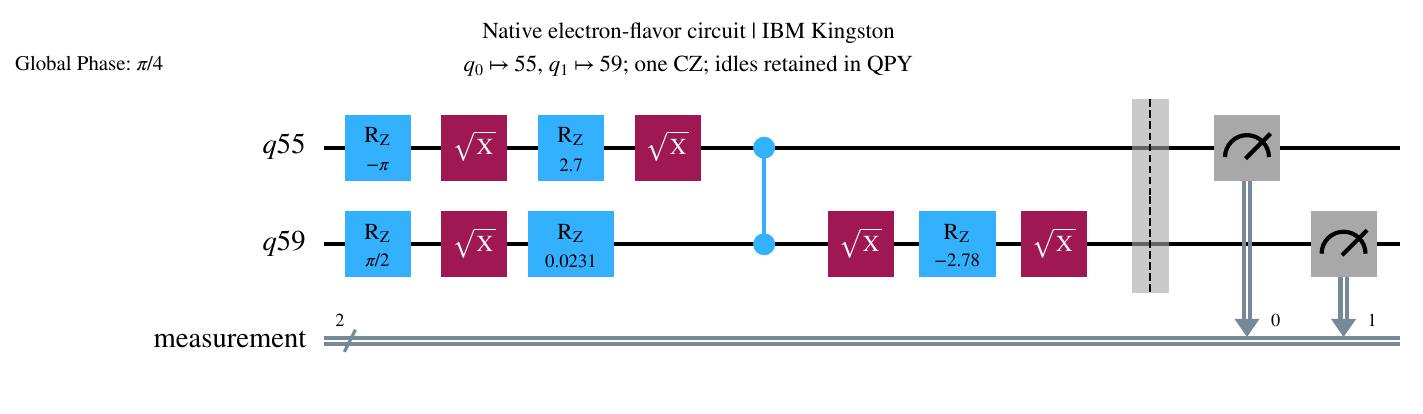}\\[1em]
\includegraphics[page=2,width=\textwidth]{native_circuits.pdf}
\caption{Native Kingston circuits fixed before execution at $\zeta=1$: electron-flavor projection (upper) and $Y$-basis measurement (lower). Wire labels identify physical qubits 55 and 59. Each circuit contains one CZ gate. The angles shown in the figure are rounded. The ancillary circuit files contain the unrounded gate parameters and scheduled delays. Barriers shown in the diagrams are omitted from the executed circuits. Terminal measurements are grouped visually
without changing the order of the native unitary gates.}
\label{fig:native}
\end{figure}

Figure~\ref{fig:native} shows the endpoint circuits compiled into
the processor's supported gates. Qiskit 2.5.2 compiles the circuits
at optimization level three with exact synthesis
\cite{Javadi-Abhari:2024kbf}. We fix the layout before sampling and retain the
physical-to-classical bit mapping in every estimator.
The native basis contains $R_z$, $\sqrt X$, $X$, and CZ gates.
The unitary $\sqrt X$ is the native square root of the Pauli-$X$ gate.
The saved Qiskit circuit files in QPY format retain scheduled delays
and full-precision angles; the figures omit idle delays for readability.

We obtain an independent classical solution by exponentiating the
matrix representation of the generator in eq.~\eqref{eq:master}.
Stacking the columns of the density matrix into a vector gives the
Liouvillian matrix on the eight-dimensional mass--energy state
space~\cite{Lindblad:1975ef,Kopp:2026tnx}:
\begin{equation}
 \widehat{\Liouv}_\zeta=
 -\ii(\mathbb 1\otimes H-H^{\mathrm T}\otimes\mathbb 1)
 +\sum_r\left[L_r^*\otimes L_r
 -\frac12\mathbb 1\otimes L_r^\dagger L_r
 -\frac12(L_r^\dagger L_r)^{\mathrm T}\otimes\mathbb 1\right].
 \label{eq:liouvillian-matrix}
\end{equation}
The index $r$ runs over the two jump operators for the transition from $E_H$ to $E_L$. The matrix exponential $\exp(L\widehat{\Liouv}_\zeta)$ is evaluated
with a scaling-and-squaring  algorithm~\cite{doi:10.1137/09074721X,Virtanen_2020}.
For the full channel, we compare every operator
$\ket{i,H}\!\bra{j,H}$, with $i,j=1,\ldots,4$.
For the compact circuit, we compare the normalized daughter block against this evolution and against the reduced five-qubit state. The recorded maximum absolute differences are \cite{Wood:2011zvw}
\begin{equation}
 \begin{array}{lr}
 \text{Full source-space channel versus Liouvillian} &1.57\times10^{-14},\\
 \text{Compact daughter state versus Liouvillian} &3.56\times10^{-16},\\
 \text{Compact versus five-qubit daughter state} &1.36\times10^{-15},\\
 \text{Compact-circuit CP identity} &6.51\times10^{-19}.
 \end{array}
 \label{eq:validation-errors}
\end{equation}
The first three are maximum matrix-entry differences, while the last is an absolute probability difference. The tests cover the two daughter pairs and five overlap values. Additional compact-state checks use $\kappa=0.1,1,3$; flavor checks include the benchmark phase, its reversal,
and zero. The common numerical tolerance is $10^{-12}$.

\begin{figure}[tbp]
\centering
\includegraphics[width=\textwidth]{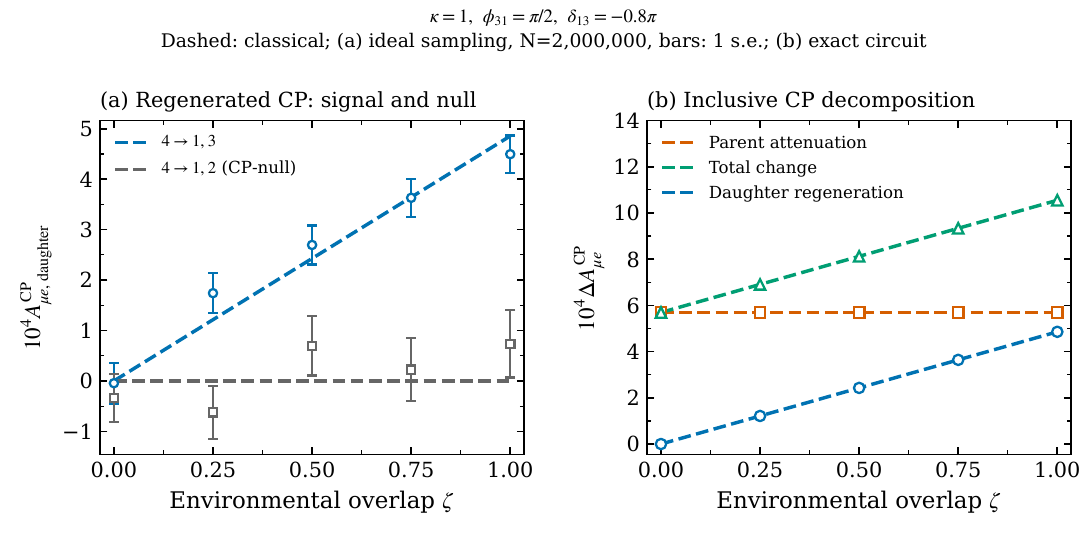}
\caption{Full five-qubit simulation. Panel (a) compares the
pair-$(1,3)$ regenerated asymmetry with the pair-$(1,2)$ control using
$2\times10^6$ ideal shots per appearance circuit.
Panel (b) contains exact statevector parent, daughter, and total
contributions to eq.~\eqref{eq:cp-decomposition}.
Dashed curves denote classical values, and error bars in panel (a)
denote one standard error. The displayed ordinates include $10^4$.}
\label{fig:full-validation}
\end{figure}

\begin{figure}[tbp]
\centering
\includegraphics[width=\textwidth]{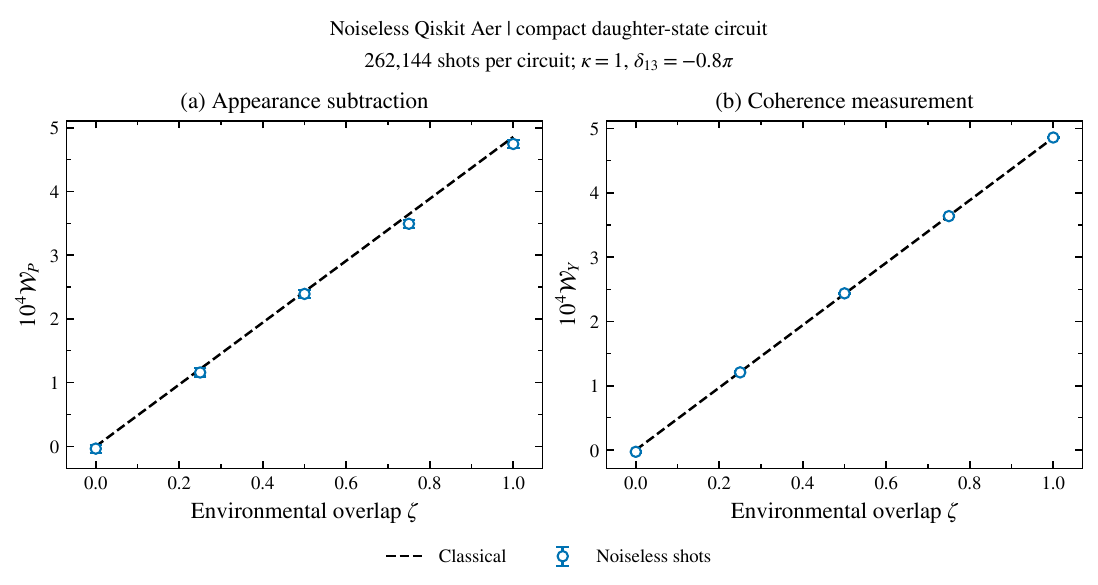}
\caption{Noiseless compact-circuit estimates from Qiskit Aer with 262,144 shots per circuit. Panels show $\mathcal W_P$ and $\mathcal W_Y$ against their dashed classical predictions. Bars denote one standard error. An independently sampled incoherent reference is shared across the appearance estimates, and the associated covariance is retained.}
\label{fig:ideal}
\end{figure}

Figure~\ref{fig:full-validation} compares the full-channel CP predictions, testing parent attenuation and daughter regeneration together before the conditional reduction. Figure~\ref{fig:ideal} then tests the compact-circuit
observables with finite-shot sampling. Both calculations use the same source and decay normalization.

Noisy simulations use Qiskit Aer with a noise model based on Kingston calibration data, including gate errors, relaxation, and dephasing during gates and idle intervals.
Using the calibration matrix $M$ defined in
eq.~\eqref{eq:assignment-definition}, we model correlated errors in
the final two-qubit measurement by the channel
\begin{equation}
 \mathcal M(\rho)=\sum_{u,v}M_{uv}
 \ket u\!\bra v\rho\ket v\!\bra u,
 \qquad K_{uv}^{\mathcal M}=\sqrt{M_{uv}}\ket u\!\bra v.
 \label{eq:noise-assignment}
\end{equation}
The full $4\times4$ matrix retains correlations between the two
recorded bits. Its column sums are unity, ensuring completeness of
the Kraus operators.

\begin{figure}[tbp]
\centering
\includegraphics[width=\textwidth]{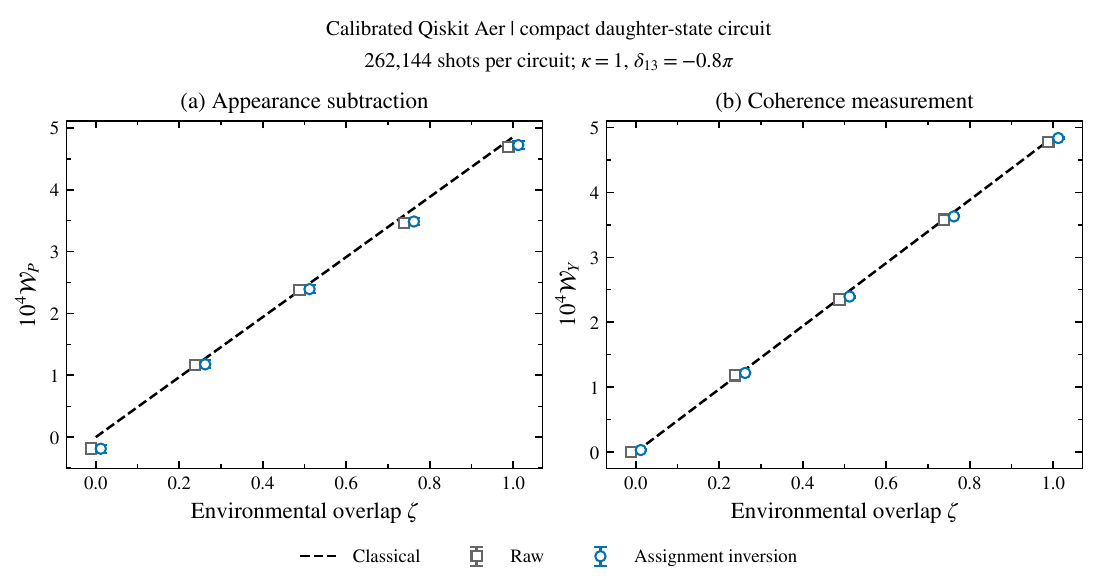}
\caption{Local Aer simulation of the native compact circuits using a calibrated noise model. Each observable-measurement circuit uses 262,144 shots, and each of the four calibration states uses 4,096 shots per simulated calibration, with two calibrations in total. Symbols distinguish raw and corrected
estimates. Error bars include uncertainties from both observable and calibration measurements; dashed curves show the classical prediction. Horizontal offsets separate the series.}
\label{fig:noisy}
\end{figure}

\begin{figure}[tbp]
\centering
\includegraphics[width=\textwidth]{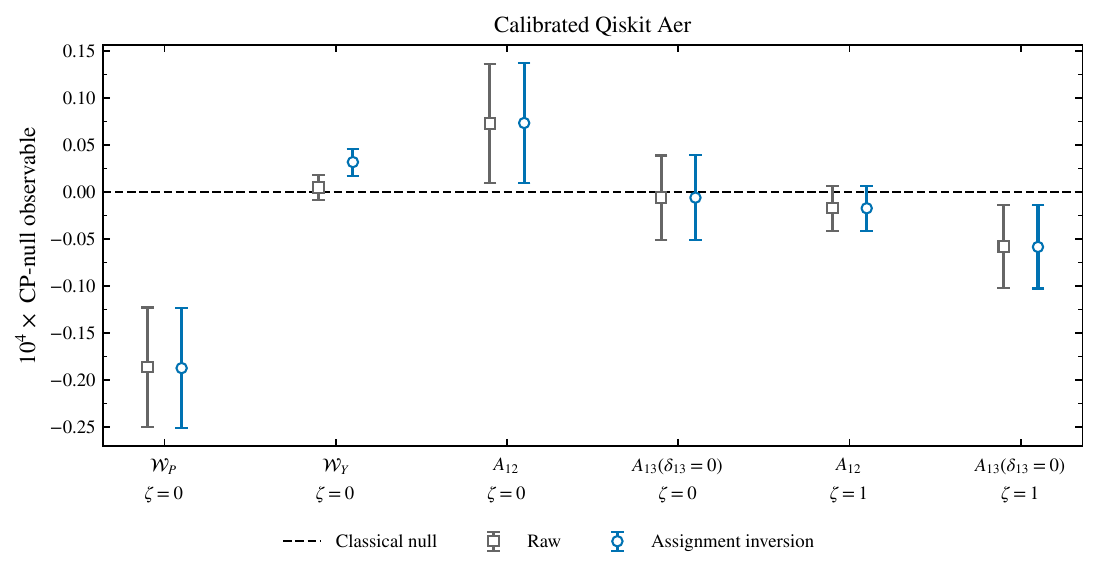}
\caption{Local noisy simulations of the null controls using the calibration model
and sampling allocations of figure~\ref{fig:noisy}.
The controls are $\mathcal W_P$ and $\mathcal W_Y$ at $\zeta=0$,
and the pair-$(1,2)$ and zero-phase appearance differences at
$\zeta=0$ and $1$.
Error bars denote one standard error, and the dashed line marks the
classical prediction of zero. Raw and corrected estimates are shown separately.}
\label{fig:noisy-nulls}
\end{figure}

Figures~\ref{fig:noisy} and~\ref{fig:noisy-nulls} display the sampled noise-model results. Exact density-matrix means separately quantify model bias. This separation is useful near a null, where a finite-shot sample can differ from zero even when the model mean is small. The figure simulations use more shots than the processor run;
table~\ref{tab:resources-main} fixes the actual hardware allocation.

\begin{figure}[tbp]
\centering
\includegraphics[width=0.90\textwidth]{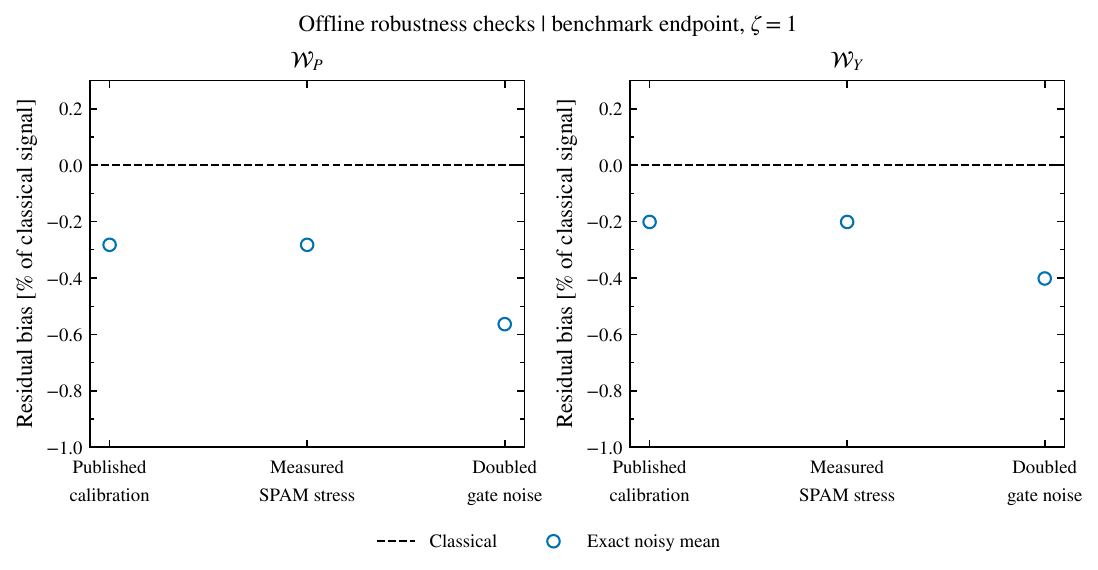}
\caption{Exact endpoint biases after measurement-error correction, relative to $S_{\CP}$. The three noise models use device calibration data, an empirical two-bit calibration matrix, and the same empirical matrix
with each gate-error channel applied twice. Symbols show density-matrix results without finite-shot sampling; the dashed line marks zero bias.}
\label{fig:robustness}
\end{figure}

We also replace the device-calibration matrix with an empirical two-bit matrix derived from an earlier five-qubit calibration, averaging over the prepared states and summing over the measured outcomes of the other three qubits. This tests sensitivity to a different measurement response. A third model uses this empirical matrix and applies each gate-error channel twice. Figure~\ref{fig:robustness} shows the endpoint biases after correction
for these three stationary noise models, whose parameters remain
fixed throughout the simulated experiment.

The model with each gate-error channel applied twice gives the largest
absolute biases: $0.563\%$ of $S_{\CP}$ for $\mathcal W_P$ and
$0.401\%$ for $\mathcal W_Y$. These simulations guided the layout
and shot allocation; stability during the processor experiment is
examined in appendix~\ref{app:calibration}.

\FloatBarrier
\section{Measurement calibration and error mitigation}
\label{app:calibration}

Errors in the recorded bit values can shift the small probability
differences used to reconstruct the regenerated CP contribution.
We correct the joint two-qubit response by linear inversion of a
calibrated measurement matrix
\cite{Maciejewski:2019jor,Bravyi:2020bpr}.

We calibrate the joint response of the daughter and environment qubits by preparing each of the four computational-basis states. The measurement calibration matrix is defined by
\begin{equation}
M_{uv}
=
P(\text{record }u\mid\text{prepare }v),
\qquad
u,v\in\{00,01,10,11\}.
\label{eq:assignment-definition}
\end{equation}
Each column gives the distribution of recorded outcomes for one intended input state. The full $4\times4$ matrix retains correlations between the two recorded bits. The calibration captures the combined effects of preparation, reset, and measurement in the basis-state circuits. Matrix inversion treats this response as an effective classical error channel.

Each input state is measured 4,096 times before the measurements of the CP observables and 4,096 times after their completion. We compare the two calibration matrices before combining their counts. The largest difference between corresponding matrix entries is $1.29$ standard errors, below the predefined threshold of $3.5$.
Pooling the two datasets then supplies 8,192 outcomes per
calibration input.

\begin{figure}[tbp]
\centering
\includegraphics[width=\textwidth]{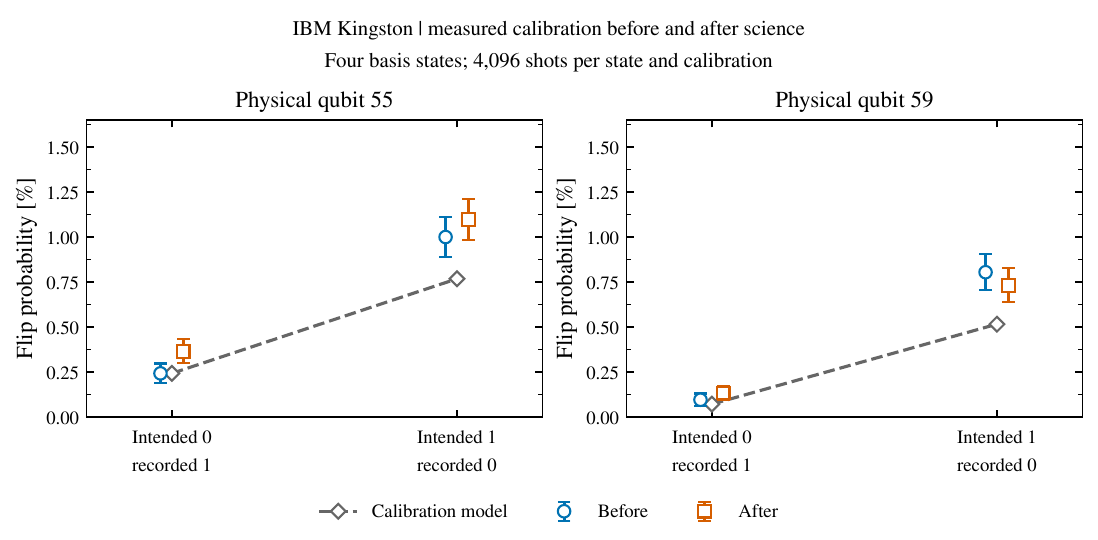}
\caption{
Probabilities of recording each bit incorrectly, measured before and after the CP-observable measurements. For each intended bit value, the probability is averaged over the two preparations of the other qubit. Each joint input state receives 4,096 repetitions per calibration run. Error bars represent one standard error, calculated from the
independent binomial counts for the two preparations of the other qubit. Grey dashed lines show the predictions of the saved noise model.
}
\label{fig:calibration}
\end{figure}

Figure~\ref{fig:calibration} shows the probability of recording each bit incorrectly. The largest measured marginal flip probability is approximately $1.10\%$. For each marginal probability, the one-sided $95\%$ Wilson upper confidence bound lies below the predefined $5\%$ threshold \cite{Wilson:1927xyh}.
The spectral condition numbers of the initial and final
calibration matrices are $1.026$ and $1.028$, respectively,
both below the acceptance threshold of $1.5$.
Here the spectral condition number is the ratio of the largest to the smallest singular value. The pooled matrix has condition number $1.027$, indicating a numerically stable inversion.

For an observed frequency vector
$\widehat{\boldsymbol f}_k$ from measurement circuit $k$,
we calculate
\begin{equation}
 \widehat{\boldsymbol q}_k
 =
 \widehat M^{-1}\widehat{\boldsymbol f}_k.
 \label{eq:assignment}
\end{equation}
Here $\widehat M$ is estimated from the pooled calibration counts,
and $\widehat{\boldsymbol q}_k$ is the vector of corrected frequency
estimates. This procedure follows the linear-inversion treatment of measurement errors
\cite{Maciejewski:2019jor,Bravyi:2020bpr}.
We apply the joint correction before summing over the
environment outcomes.

We construct the CP estimators as linear combinations of the corrected frequencies. The matrix inversion is unconstrained, and we retain negative components that can arise from finite sampling. Clipping these components would alter the estimators and could bias small probability differences near a null. The calibration correction
acts on the measurement statistics and leaves the physical prediction unchanged.

The CP-observable measurements were divided into two blocks
with independently shuffled circuit order.
Repeating the settings permits a comparison of the reconstructed observables at different times during the run.
The largest difference between the two blocks is $2.05$
standard errors, including the covariance from their common
calibration. Both the calibration comparison and the block comparison remain below the predefined $3.5$-standard-error threshold.

Appendix~\ref{app:validation} reports the biases predicted by the
stationary noise models. The before-and-after calibrations and
shuffled measurement blocks instead test stability during the
processor experiment. The two CP estimators also use distinct
measurement rotations, providing a consistency check through
different circuits.
\FloatBarrier
\section{Statistical uncertainties}
\label{app:statistics}

The quoted uncertainties include statistical fluctuations in both the observable measurements and the calibration counts.
We retain correlations arising from the shared calibration matrix and incoherent reference, while keeping all physical parameters and normalization weights fixed at their benchmark values.

For circuit $k$, let $\widehat{\boldsymbol f}_k$ denote the
observed frequency vector from $N_k$ repetitions.
The corrected estimator for observable $O$ is
\begin{equation}
 \widehat O
 =
 \sum_k \boldsymbol w_{O,k}^{\mathrm T}
 \widehat M^{-1}\widehat{\boldsymbol f}_k,
 \label{eq:weighted-estimator}
\end{equation}
where $\widehat M$ is the estimated calibration matrix and
$\boldsymbol w_{O,k}$ contains the outcome weights, subtraction signs, and physical normalization factors defined in section~\ref{sec:quantum}.

For a frequency vector $\boldsymbol f$ obtained from $N$
repetitions, we estimate its multinomial covariance as
\begin{equation}
 C(\boldsymbol f,N)
 =
 \frac{\operatorname{diag}(\boldsymbol f)
       -\boldsymbol f\boldsymbol f^{\mathrm T}}{N-1},
 \label{eq:multinomial}
\end{equation}
where $\operatorname{diag}(\boldsymbol f)$ places the components of $\boldsymbol f$ on the diagonal.
Thus
$C_k=C(\widehat{\boldsymbol f}_k,N_k)$
for the observable measurements and
$C_v^{\mathrm{cal}}=C(\widehat{\boldsymbol m}_v,8192)$
for calibration input $v$, where
$\widehat{\boldsymbol m}_v$ is column $v$ of $\widehat M$.

We propagate these covariances by linearizing the corrected estimators
with respect to the measured frequencies and calibration-matrix
entries~\cite{7920614895714e3eb6745abd6cf33338}.
We define
$\boldsymbol a_{O,k}=\widehat M^{-\mathrm T}\boldsymbol w_{O,k}$,
$\widehat{\boldsymbol q}_k=\widehat M^{-1}\widehat{\boldsymbol f}_k$,
and
$G_O=-\sum_k\boldsymbol a_{O,k}
\widehat{\boldsymbol q}_k^{\mathrm T}$.
The covariance of two reconstructed observables is
\begin{equation}
 \begin{aligned}
 \operatorname{Cov}(\widehat O,\widehat O')
 &=
 \sum_{k\in O\cap O'}
 \boldsymbol a_{O,k}^{\mathrm T}C_k\boldsymbol a_{O',k}
 \\
 &\quad+
 \sum_v (G_O)_{:v}^{\mathrm T}
 C_v^{\mathrm{cal}}(G_{O'})_{:v}.
 \end{aligned}
 \label{eq:full-covariance}
\end{equation}
The first sum includes shared measurement datasets; the second accounts for the common calibration. The notation $(G_O)_{:v}$ denotes column $v$. Each error bar represents one standard error, $\sigma_O=\sqrt{\operatorname{Var}(\widehat O)}$. For the difference between the two CP observables,
\begin{equation}
 \sigma_{P-Y}^{2}
 =
 \operatorname{Var}(\widehat{\mathcal W}_P)
 +
 \operatorname{Var}(\widehat{\mathcal W}_Y)
 -
 2\operatorname{Cov}
 (\widehat{\mathcal W}_P,\widehat{\mathcal W}_Y).
 \label{eq:readout-difference-variance}
\end{equation}

A multinomial bootstrap with 5,000 replicates checks the
analytic propagation \cite{Efron:1979bxm}.
Each replicate resamples the measurement histograms and
calibration columns at their original count totals, repeats
the inversion, and uses the same resampled calibration and
reference data across all observables.
The bootstrap standard errors agree with the analytic values
to within $1.2\%$.

The acceptance criteria were fixed before hardware execution, using
the benchmark signal $S_{\CP}$ defined in eq.~\eqref{eq:b2-total}.
Each primary estimator was required to satisfy
\begin{equation}
 \sigma_{\mathcal W}<0.10S_{\mathrm{CP}},
 \qquad
 |\widehat{\mathcal W}-S_{\mathrm{CP}}|
 +1.96\sigma_{\mathcal W}<0.15S_{\mathrm{CP}},
 \label{eq:signal-acceptance}
\end{equation}
with mutual agreement within $3.5$ standard errors.
For each null estimator $\widehat A$, the criterion was
\begin{equation}
 |\widehat A|+1.96\sigma_A<0.20S_{\mathrm{CP}}.
 \label{eq:null-acceptance}
\end{equation}
This bound tests whether the residual is small relative to
the benchmark signal, including its statistical uncertainty.
All six predefined controls satisfy it, with bounds between
$2.3\%$ and $16.8\%$ of $S_{\mathrm{CP}}$.

\begin{table}[tbp]
\centering
\small
\begin{tabular}{@{}lccc@{}}
\toprule
Control & $\zeta$ & Corrected result $\times10^4$
& Bound$/S_{\mathrm{CP}}$ \\
\midrule
$\mathcal W_P$ & 0 & $0.3428\pm0.2215$ & $16.0\%$ \\
$\mathcal W_Y$ & 0 & $-0.0063\pm0.0535$ & $2.3\%$ \\
Pair $(1,2)$ & 0 & $0.3117\pm0.2565$ & $16.8\%$ \\
$\delta_{13}=0$ & 0 & $0.3138\pm0.1809$ & $13.8\%$ \\
Pair $(1,2)$ & 1 & $0.1474\pm0.1051$ & $7.3\%$ \\
$\delta_{13}=0$ & 1 & $-0.1020\pm0.1776$ & $9.3\%$ \\
\midrule
Direct pair $(1,3)$ diagnostic
& 0 & $0.3556\pm0.1809$ & $14.6\%$ \\
\bottomrule
\end{tabular}
\caption{
Controls after measurement-error correction, with uncertainties of one standard error. The final column lists
$(|\widehat A|+1.96\sigma_A)/S_{\mathrm{CP}}$.
The first six rows form the predefined control set.
The final row records the direct appearance difference without incoherent-reference subtraction and shares counts with
$\mathcal W_P$.
}
\label{tab:nulls}
\end{table}

The additional direct pair-$(1,3)$ diagnostic at $\zeta=0$
lies $1.97$ standard errors from zero.
It is reported separately from the predefined controls,
with its shared-count correlations retained in the analysis.

\FloatBarrier
\bibliographystyle{JHEP}
\bibliography{references}
\end{document}